\documentclass[preprint,aps,amsmath,amssymb,superscriptaddress,floatfix]{revtex4-1}

\usepackage{graphicx}
\usepackage{dcolumn}
\usepackage{bm}

\usepackage{tabularx} 
                                                                                                                                                                                                    
\usepackage{subfigure}
\usepackage{epsfig}
\newcommand{\no}{\nonumber}

\renewcommand{\i}{{\text{i}}}

\usepackage{color}
\usepackage{txfonts}

\graphicspath{{../PhD_Thesis/UoMThesis/figures/Baby/}{Plot_mat/}{freq_crit/}}

\begin{document}

\title{Isospinning baby Skyrmion solutions}

\author{Richard A.~Battye}
\email{richard.battye@manchester.ac.uk}
\affiliation{%
Jodrell Bank Centre for Astrophysics, University of Manchester, Manchester M13 9PL, U.K.\\
}%
\author{Mareike Haberichter}%
 \email{m.haberichter@kent.ac.uk}
\affiliation{%
Jodrell Bank Centre for Astrophysics, University of Manchester, Manchester M13 9PL, U.K.\\
}%
\affiliation{%
School of Mathematics, Statistics and Actuarial Science,\\ University of Kent, Canterbury CT2 7NF, U.K.\\
}%

\date{\today}

\begin{abstract}
We perform full two-dimensional (2D) numerical relaxations of isospinning soliton solutions in the baby Skyrme model in which the global $O(3)$ symmetry is broken by the 2D analogue of the pion mass term in the Skyrme model. In our calculations we explicitely allow the isospinning solitons to deform and to break the symmetries of the static configurations. We find that stable isospinning baby Skyrme solutions can be constructed numerically for all angular frequencies $\omega\le \text{min}(\mu,1)$, where $\mu$ is the mass parameter of the model. Stable, rotationally-symmetric baby Skyrmion solutions for higher angular velocities are simply an artefact of the hedgehog approximation. Isospinning multisoliton solutions of topological charge $B$ turn out to be unstable to break up into their $B$ charge-1 constituents at some critical breakup frequency value. Furthermore, we find that for $\mu$ sufficiently large the rotational symmetry of charge-2 baby Skyrmions becomes broken at a critical angular frequency $\omega$.

\end{abstract}

\pacs{Valid PACS appear here}
\maketitle

\section{Introduction}

Baby Skyrmions \cite{Piette1995,Piette:1994jt} are topological soliton solutions in $(2+1)$-dimensional versions of the  Skyrme model \cite{Skyrme:1961vq} for nuclear physics. They can be used to describe quasiparticle excitations in ferromagnetic quantum Hall systems \cite{Sondhi:1993zz,Walet:2001zz} or can arise as stable skyrmion spin textures in various other condensed-matter systems such as helical magnets $\text{Fe}_{1-x}\text{CO}_x\text{Si}$ \cite{Onose:2010} and $\text{MnSi}$ \cite{Muhlbauer:2009}.

In this article, our numerical study of isospinning soliton solutions in baby Skyrme models is not only motivated by their physical relevance but even more by the similiarities to their analogues in $3+1$ dimensions, which are known as Skyrmions. Using the baby Skyrme model as a simplified model to guide investigations in the full $(3+1)$-dimensional Skyrme theory has proven to be useful in the study of various Skyrme soliton properties. For example, scattering processes of baby Skyrmions  \cite{Peyrard:1990hc,Piette:1994mh} strongly ressemble those of Skyrmions \cite{Schroers:1993yk} and allow us to study the long-range forces between static, moving and spinning Skyrmions with moderate numerical effort.

Suitably quantized Skyrmions \cite{Adkins:1983ya,Manko:2007pr} of topological charge $B$ are promising candidates to model spin and isospin states of nuclei with baryon number $B$. However, most attempts have so far been either based on a rigid-body type approximation \cite{Adkins:1983ya,Manko:2007pr} -- the soliton's shape is taken to be rotation frequency independent -- or only considered axially symmetric deformations \cite{Battye:2005nx,Houghton:2005iu,Fortier:2008yj} of the spinnning Skyrme configurations. As pointed out by several authors \cite{Braaten:1984qe,Battye:2005nx,Houghton:2005iu}, neglecting any deformations that could arise from the dynamical terms in the Skyrme model is far from an adequate approximation, and working within an axially symmetric solution ansatz is only a valid simplification for few Skyrmions since most Skyrme solitons are not axially symmetric.

In this article, we use the baby Skyrme model as testing ground to numerically investigate how the Skyrme soliton's geometrical shape, its mass and its moment of inertia change when classically isorotating around the $z$ axis. 

To our knowledge, up to now there only exists research on isospinning, rotationally symmetric charge-1 and charge-2 soliton solutions of the conventional baby Skyrme model \cite{Piette:1994mh,Acus:2009df} and of two modified baby Skyrme models where the domain $\mathbb{R}^2$ is replaced by a 2-sphere and by a unit disk, respectively \cite{Hen:2007in,Hen:2008ha,Karliner:2009at}. Furthermore, isospinning rotationally invariant solitons have been constructed analytically \cite{Gisiger:1996vb} in the Bogomol‘nyi-Prasad-Sommerfield (BPS) limit of the conventional baby Skyrme model. Compared to previous work, we do not impose any symmetries on the isopinning Skyrme configurations and we do not apply the rigid-body approximation in our 2D numerical computations. Our calculations are performed for isospinning  soliton solutions up to charge 6 in the baby Skyrme model with a potential analogous to that used in the $(3+1)$-dimensional Skyrme model.

This article is structured as follows. In Sec.~\ref{Sec_Baby} we briefly review the baby Skyrme model and explain how isospinning soliton solutions arise there. Then, in Sec.~\ref{Sec_Baby_IC} we create suitable initial field configurations of nonzero baryon number $B$, which in Sec.~\ref{Sec_Baby_Static} are numerically minimized using a 2D gradient flow evolution algorithm. The resulting baby Skyrme configurations of vanishing isospin serve us as initial fields for our numerical simulations of isospinning baby Skyrmions in Sec.~\ref{Sec_Baby_Spin_Res}. We present our conclusions in Sec.~\ref{Sec_Baby_Con}.  

\section{Classically Isospinning Baby Skyrmions}\label{Sec_Baby}

The Lagrangian density of the $(2+1)$-dimensional baby Skyrme model is defined by
\begin{align}
\mathcal{L}&=\frac{1}{2}\partial_\mu\boldsymbol\phi\cdot\partial^\mu\boldsymbol\phi-\frac{1}{4}\left(\partial_\mu\boldsymbol{\phi}\times\partial_\nu\boldsymbol{\phi}\right)^2-V(\boldsymbol{\phi})\,,
\label{Lag_Baby}
\end{align}
where $\boldsymbol{\phi}=(\phi_1,\phi_2,\phi_3)$ is a real vector field of unit length and hence the target space of $\boldsymbol{\phi}$ is given by a 2-sphere $S^2_{\text{iso}}$. The first term in (\ref{Lag_Baby}) is known as the $O(3)$ sigma model term, the second is the $(2+1)$-dimensional analogue of the Skyrme term, and the last term is a potential term which is needed to stabilize the soliton's size. 

Finite energy solutions in model (\ref{Lag_Baby}) require that the vector field $\boldsymbol{\phi}$ has to approach a constant value $\boldsymbol{\phi}_\infty$ at spatial infinity, i.e. $\boldsymbol{\phi}(\boldsymbol{x},t)\rightarrow\boldsymbol{\phi}_\infty$ as $|\boldsymbol{x}|\rightarrow\infty$ for all time $t$. Consequently, this boundary condition results in a one-point compactification of physical space $\mathbb{R}^2$ into a 2-sphere $S^2_{\text{space}}$, and hence a field configuration $\boldsymbol{\phi}$ can be labeled by the winding number $B$ of the map $S^2_\text{space}\mapsto S^2_{\text{iso}}$, explicitly given by
\begin{align}
B&=\text{deg}[\boldsymbol{\phi}]=\frac{1}{4\pi}\int\,\boldsymbol{\phi}\cdot\partial_1\boldsymbol{\phi}\times\partial_2\boldsymbol{\phi}\,\text{d}^2x\,.
\label{Baby_charge}
\end{align}
The topological charge $B\in\pi_2\left(S^2\right)=\mathbb{Z}$ is called baryon number in analogy to the Skyrme model \cite{Skyrme:1961vq}, and field configurations which minimize the static energy functional in a given topological sector $B$  are known as baby Skyrmions. In this article, the static energy $M_B$ of a baby Skyrmion solution of topological charge $B$ will be given in units of $4\pi B$. Our  normalization choice is motivated by the Bogomolny lower energy bound
\begin{align}
M_B\geq4\pi B\,,
\label{Bogo_Lump}
\end{align}
and simplifies the comparison with the energy values stated in the literature (e.g. \cite{Piette:1994mh}). Note that Eq.~(\ref{Bogo_Lump}) is the topological bound related to the $O(3)$ sigma model term in (\ref{Lag_Baby}) alone. Both the $O(3)$ sigma model term and the Skyrme and potential terms together provide an even tighter energy  bound \cite{deInnocentis:2001ur,Adam:2010jr,Speight:2010sy}.

The shape and the behaviour of baby Skyrmion solutions depend strongly on the choice of the potential term $V(\boldsymbol{\phi})$ in (\ref{Lag_Baby}). Here we will consider the $O(3)$ symmetry breaking potential term
\begin{align}
V\left(\boldsymbol{\phi}\right)&=\mu^2\left(1-\phi_3\right)\,,
\label{Lag_pot}
\end{align}
where $\mu$ is a rescaled mass parameter. Equation (\ref{Lag_pot}) is the most common potential choice,  which was originally included in \cite{Piette:1994ug,Piette:1994mh} in analogy to the pion mass term of the full $(3+1)$-dimensional Skyrme model. To ensure a finite-energy configuration the field has to approach its vacuum value $\phi_3=+1$ at spatial infinity. Choosing (\ref{Lag_pot}) in (\ref{Lag_Baby}) results in two massive modes ($\phi_1,\,\phi_2$) of mass $\mu$ and one massless ($\phi_3$). Here, the mass parameter $\mu$ is  usually chosen to be $\mu=\sqrt{0.1}\approx 0.316$ \cite{Piette:1994ug,Piette:1994mh}, so that the size of the $B=1$ soliton solution is approximately of order of the Compton wavelength of the ``mesons'' in our model. This specific choice of mass parameter was originally motivated by the full $(3+1)$-dimensional Skyrme model, in which the 1-soliton size is approximately equal to the pion's wavelength. Static multisoliton solutions in model (\ref{Lag_Baby}) with potential term (\ref{Lag_pot}) have been studied in \cite{Piette:1994ug,Eslami:2000tj,Foster:2009vk}.

To find isospinning baby Skymion solutions we perform a time-dependent $S\!O(2)$ isorotation on a static baby Skyrmion configuration $\boldsymbol{\phi}$ via
\begin{align}
\left(\phi_1,\,\phi_2,\,\phi_3\right)\mapsto\left(\cos\left(\alpha+\omega t\right)\phi_1+\sin\left(\alpha+\omega t\right)\phi_2,\,-\sin\left(\alpha+\omega t\right)\phi_1+\cos\left(\alpha+\omega t\right)\phi_2,\,\phi_3\right)\,,
\label{Ansatz_spin}
\end{align} 
where the rotation axis is chosen to be $(0,0,1)$, $\omega$ is the angular frequency, and the angle $\alpha\in[0,2\pi)$. Substituting the dynamical ansatz (\ref{Ansatz_spin}) into the Lagrangian (\ref{Lag_Baby}) gives
\begin{align}
L=\frac{1}{2}\Lambda_B\omega^2-M_B\,,
\end{align}
where $M_B$ is the classical soliton mass
\begin{align}
M_B&=\int\left\{\frac{1}{2}\partial_i\boldsymbol{\phi}\cdot\partial_i\boldsymbol{\phi}+\frac{1}{4}\left[\left(\partial_i\boldsymbol{\phi}\cdot\partial_i\boldsymbol{\phi}\right)^2-\left(\partial_i\boldsymbol{\phi}\cdot\partial_j\boldsymbol{\phi}\right)\left(\partial_i\boldsymbol{\phi}\cdot\partial_j\boldsymbol{\phi}\right)\right]+V\left(\boldsymbol{\phi}\right)\right\}\,\text{d}^2x\,,
\label{baby_mass}
\end{align}
and $\Lambda_B$ is the moment of inertia
\begin{align}
\Lambda_B&=\int\Bigg\{\left(\phi_1^2+\phi_2^2\right)\left(1+\partial_k\boldsymbol\phi\cdot\partial_k\boldsymbol\phi\right)-\left(\boldsymbol\phi\times\partial_k\boldsymbol\phi\right)_3\left(\boldsymbol\phi\times\partial_k\boldsymbol\phi\right)_3\!\Bigg\}\, \text{d}^2x\,.
\label{Inertia_U33}
\end{align}
The Noether charge associated with the $S\!O(2)$ transformation (\ref{Ansatz_spin}) is the conserved total isospin $K=\omega\Lambda$. 

As shown in \cite{Harland:2013uk} the problem of constructing isospinning soliton solutions in Skyrme models can be formulated in terms of the following two variational problems for $\boldsymbol{\phi}$:
\begin{itemize}
\item[(1)] Extremize the pseudoenergy functional $F_\omega(\boldsymbol{\phi})=-L$ for fixed $\omega$\,,
\item[(2)] Extremize the total energy functional $H=M_B+K^2/(2\Lambda_B)$ for fixed $K$.
\end{itemize}
We performed most of our numerical relaxations for both methods and verified that we obtained the same soliton shape and dependence of the soliton's energy on the rotation frequency $\omega$.

Note that the  pseudoenergy functional $F_\omega\left(\boldsymbol{\phi}\right)=M_B-\frac{1}{2}\Lambda_B\omega^2$ takes the form
\begin{align}\label{Baby_Pseudo_energy_phi}
F_\omega(\boldsymbol{\phi})&=\int\left[\frac{1}{2}\Bigg\{\left[1-\omega^2\left(1-\phi_3^2\right)\right]\left(\partial_i\boldsymbol{\phi}\cdot\partial_i\boldsymbol{\phi}\right)+\omega^2\left(\boldsymbol{\phi}\times\partial_i\boldsymbol{\phi}\right)_3\left(\boldsymbol{\phi}\times\partial_i\boldsymbol{\phi}\right)_3\Bigg\}\no\right.\\
&\left.+\frac{1}{4}\left(\partial_i\boldsymbol{\phi}\times\partial_j\boldsymbol{\phi}\right)^2+V_\omega(\boldsymbol{\phi})\right]\,\text{d}^2x\,,
\end{align}
where the effective, deformed  potential $V_\omega(\boldsymbol{\phi})$ is given by
\begin{align}
V_\omega(\boldsymbol{\phi})&=V(\boldsymbol{\phi})-\frac{\omega^2}{2}\left(1-\phi_3^2\right)\,.
\label{Baby_pseudo_poteff}
\end{align}
It was pointed out in \cite{Harland:2013uk} that isospinning soliton solutions in Skyrme-like models suffer from two different types of instabilities: One is due to the nullification of the terms quadratic in first spatial derivatives in (\ref{Baby_Pseudo_energy_phi}) at some critical frequency value $\omega_1$, and the other is related to the vanishing of the effective potential term $V_\omega(\boldsymbol{\phi})$ (\ref{Baby_pseudo_poteff}) at a second critical frequency value $\omega_2$. Recall \cite{Harland:2013uk}  that for $0<\omega<1$ the terms in the curly brackets in (\ref{Baby_Pseudo_energy_phi}) effectively describe the geometry  of a squashed sphere $S^2$ deformed along the direction $\boldsymbol{\phi}_\infty=(0,0,1)$. For $\omega>\omega_1=1$, the metric becomes singular and the pseudoenergy of $\boldsymbol{\phi}$ is no longer bounded from below. The second critical frequency $\omega_2$ is sensitive to the concrete potential choice $V(\boldsymbol{\phi})$ in (\ref{Lag_Baby}) and follows from the condition that the deformed potential term (\ref{Baby_pseudo_poteff}) has to be positive and nonzero in order to allow for stable isospinning soliton solutions. Consequently, stable isospinning soliton solutions can only be constructed for all angular frequencies $\omega\le \text{min}\left\{1,\omega_2\right\}$. For the potential (\ref{Lag_pot}) the second critical value is given by $\omega_2=\mu$, the meson mass of the model.

\section{Initial Conditions}\label{Sec_Baby_IC}

We create suitable initial field configurations with nontrivial baryon number $B$ by linear superposition of static $B=1$ hedgehog solutions. These initial baby Skyrmion fields will then be used as input for a 2D gradient flow code to search for static ($\omega=0$) minimal-energy soliton solutions in model (\ref{Lag_Baby}) with the standard potential term (\ref{Lag_pot}).  

Hedgehog fields -- fields for which a spatial rotation can be compensated by an isospin rotation -- are of the form
\begin{align}
\boldsymbol{\phi}(\boldsymbol{x})&=\left(\sin f \cos\left(B\theta-\chi\right),\,\sin f\sin\left(B\theta-\chi\right),\cos f\right)\,,
\label{Baby_hedge}
\end{align}
where $(r,\theta)$ are polar coordinates in the plane, $\chi\in[0,2\pi)$ is a phase shift, and $f(r)$ is a monotonically decreasing radial profile function with boundary conditions $f(0)=\pi$ and $f(\infty)=0$. Substituting (\ref{Baby_hedge}) in the energy functional (\ref{baby_mass})  ($\mu=\sqrt{0.1}$) and solving the associated Euler-Lagrange equation numerically \cite{Ascher:1979}, we find for the $B=1$ soliton mass $1.562\times4\pi$ and the corresponding moment of inertia (\ref{Inertia_U33}) is given by $7.533\times2\pi$. Here, the ``units'' are chosen to simplify comparison with the values in the literature, in particular, with  the hedgehog solutions calculated in Ref.~\cite{Piette:1994mh}. 

To construct multisoliton solutions in the baby Skyrme model it is convenient to parametrize the real three-component unit vector $\boldsymbol{\phi}$ in terms of a single complex scalar field $W$ \cite{Piette1995} via the stereographic projection 
\begin{align}
\boldsymbol{\phi}&=\frac{1}{1+|W|^2}\left(W+\overline{W},\i\left(W-\overline{W}\right),\left(1-|W|^2\right)\right)\,.
\end{align}
A  rotationally symmetric, complex field $W^{(n)}$ is given by
\begin{align}
W^{(n)}(r,\theta)=\tan\left(\frac{f}{2}\right)\exp(-\i n \theta)\,,
\label{W_field}
\end{align}
where $f$ is the solution of the reduced equation for a charge-$n$ baby Skyrmion.

Initial field configurations of baryon number $B=n$ are obtained within this complex field formalism by a linear superposition of $n$ 1-soliton solutions \cite{Eslami:2000tj} 
\begin{align}
W(x,y)=\sum_{c=1}^{n} W^{(1)}(x-x_c,y-y_c)\exp(\i\chi_c)\,,
\end{align}
where $W^{(1)}$ is the complex field of the $c$th baby Skyrmion,$(x_c,\,y_c)$ are the cartesian coordinates of the center of the $c$th baby Skyrmion, and $\chi_c$ are the $c$th baby Skyrmion's respective phase. For most of our numerical simulations we use a circular initial setup \cite{Eslami:2000tj} of $n$ equally spaced baby 1-Skyrmions with relative phase shifts $\delta \chi=\frac{2\pi}{n}$ for maximal attraction \cite{Piette:1994ug}. 

\section{Static Baby Skyrmion Solutions}\label{Sec_Baby_Static}

To find the stationary points of the energy functional $M_B$, given in (\ref{baby_mass}), we solve the associated gradient flow equation numerically. The gradient flow equation is a first-order equation in a fictious time and is obtained by setting the velocity of the field equal to minus the variation of the energy functional
\begin{align}
\dot{\boldsymbol{\phi}}=-\frac{\delta M_B}{\delta\boldsymbol{\phi}}-\lambda\boldsymbol{\phi}\,,
\label{Flow_eq}
\end{align}
where the Lagrange multiplier $\lambda$ imposes the unit vector constraint $\boldsymbol{\phi}\cdot\boldsymbol{\phi}=1$. The initial configurations are evolved according to the flow equations (\ref{Flow_eq}) on rectangular grids typically containing $(601)^2$ lattice points and with a lattice spacing $\Delta x=0.2$. Only our relaxation calculations on 1-baby skyrmion solutions are performed on finer grids with $\Delta x=0.15$ and $(401)^2$ grid points. The gradient flow equations are discretized using second-order accurate finite difference approximations for the spatial derivatives and first-order ones for the time derivatives with $\Delta t=0.005$. The Lagrange multiplier $\lambda$ is explicitely calculated at each timestep of the gradient flow evolution. Recall that we choose in this section the mass parameter $\mu=\sqrt{0.1}$ for the baby Skyrme potential (\ref{Lag_pot}).

\begin{figure}[!htb]
\centering
\includegraphics[totalheight=6.0cm]{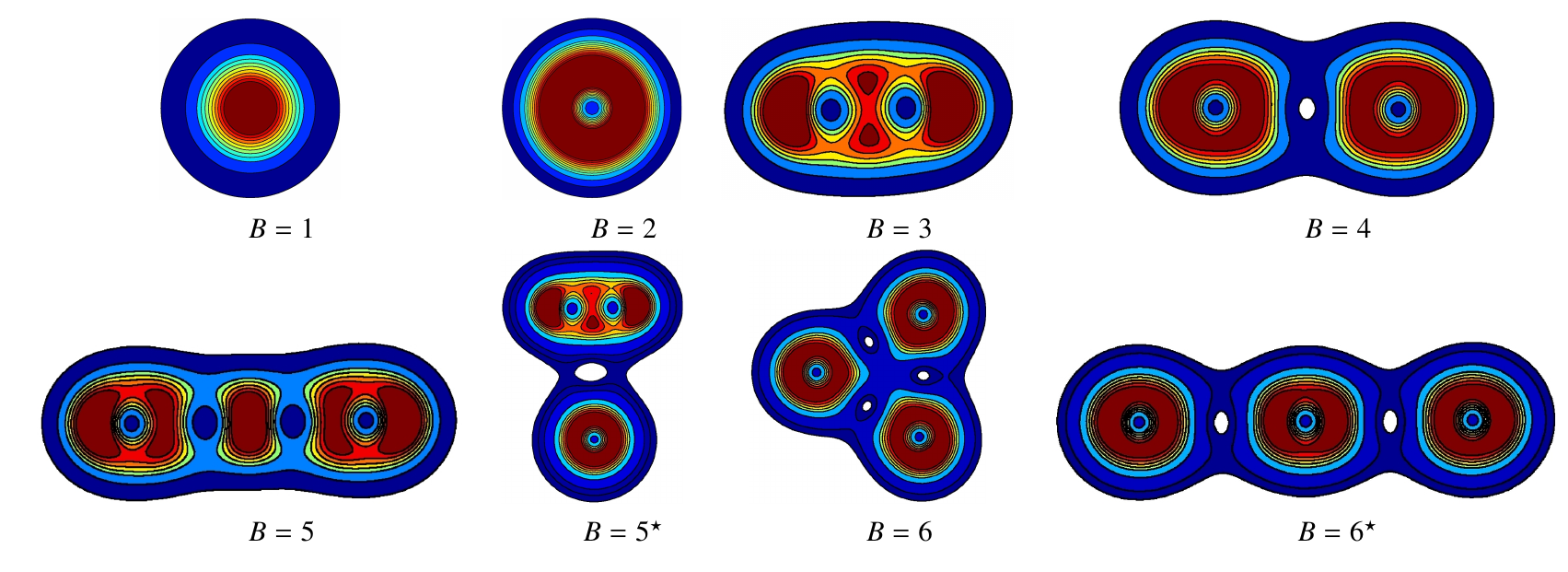}
\caption{Energy density contour plots (not to scale) for multisolitons in the conventional baby Skyrme model (\ref{Lag_pot}) with topological charges $B=1-6$. The numerical calculations were performed with the mass parameter $\mu$ set to $\sqrt{0.1}$.}
\label{Fig_Dens_VI}
\end{figure}

We list in Table~\ref{Multi_sky_V1} the energy and moment of inertia values which we obtained for static, minimum-energy configurations in the conventional baby Skyrme model. The associated energy density contour plots are displayed in  Fig.~\ref{Fig_Dens_VI}. 

\begin{table}[h!]
\caption{Multi-Skyrmion solutions of the conventional baby Skyrme model. For comparison, we include the normalized energies $M_B^\text{Foster}/B$ given in \cite{Foster:2009vk}.}
\begin{ruledtabular}
\begin{tabular}{cccccc}
 $B$& $M_B/4\pi$  &$M_B/4\pi B$&$M_B^\text{Foster}/4\pi B$ &$\Lambda_B/2\pi$ &Shape\\
\hline
1& $1.564$ & 1.564 &1.564 & 7.556        &hedgehog\\
2& $2.935$ & 1.467 & 1.468 &10.617      &one pair \\
3& $4.423$  & 1.474& 1.474 &15.389  &one triple\\
4& $5.858$ & 1.464 & 1.464& 20.524  &two pairs \\
5&$7.323$&1.464& 1.464 & 25.303    &5-chain\\
$5^\star$ &7.363&1.472     & 1.470  & 25.917 &triple+pair\\
6 &$8.778$&1.463  & 1.462& 30.742  &three pairs\\
$6^\star$& 8.786    &1.464  & 1.462     & 31.716   &6-chain\\
\end{tabular}
\end{ruledtabular}
\label{Multi_sky_V1}
\end{table}

In this model (\ref{Lag_pot}) our 2D gradient flow algorithm reproduces the axially symmetric $B=1$ and $B=2$ soliton solutions \cite{Piette:1994ug}. Their minimal energies are found to be $M_1=1.564\times4\pi$ and $M_2=2.935\times4\pi$, which is in excellent agreement with the values in the  literature \cite{Piette:1994ug,Eslami:2000tj,Foster:2009vk}. The corresponding moments of inertia are given by $\Lambda_1=7.556\times2\pi$ and $\Lambda_2=10.617\times2\pi$, respectively. For comparison, minimizing the energy functional $M_B$ within the axially symmetric ansatz (\ref{Baby_hedge}) results, for the 1-soliton, in $M_1=1.562\times 4\pi$ and $\Lambda_1=7.533\times2\pi$. The analogous calculation for $B=2$ gives an energy of $M_2=2.934\times4\pi$ and a moment of inertia $\Lambda_2=10.593\times2\pi$. As already observed \cite{Battye:2009ad} for soliton solutions in the full $(3+1)$-dimensional Skyrme model, the numerical values for the moments of inertia are a lot less accurate than the ones for the solitons' masses.

However, hedgehog solutions for topological charges $B>2$ have been shown \cite{Piette:1994ug} to be unstable against axial perturbations. The 3-baby Skyrmion configuration forms a chain of three aligned $B=1$ baby Skyrmion with a phase shift $\delta\chi=\pi$ between neighboring 1-Skyrmions. Similarly, the energy density distribution for the $B=4$ soliton is linear, but made up of two radially symmetric 2-Skyrmions. Our obtained energies $M_3=4.423\times4\pi$ and $M_4=5.858\times4\pi$ are within $0.005\%$ agreement with the energy values given in \cite{Foster:2009vk}. Relaxing a circular set-up of five $B=1$ baby Skyrmions, we reproduce the 5-chain solution \cite{Foster:2009vk}, which we confirm to be of lower energy than the $2+3$ configuration (labeled $5^\star$) \cite{Piette:1994ug}. For $B=6$ we find two configurations which can be seen as energy degenerate within our numerical accuracy: the 6-chain solution (labeled $6^\star$) \cite{Foster:2009vk} and a configuration with three 2-baby Skyrmions placed at the vertices of an equilateral triangle \cite{Piette:1994ug}. 

\begin{figure}[!htb]
\centering
\includegraphics[totalheight=6cm]{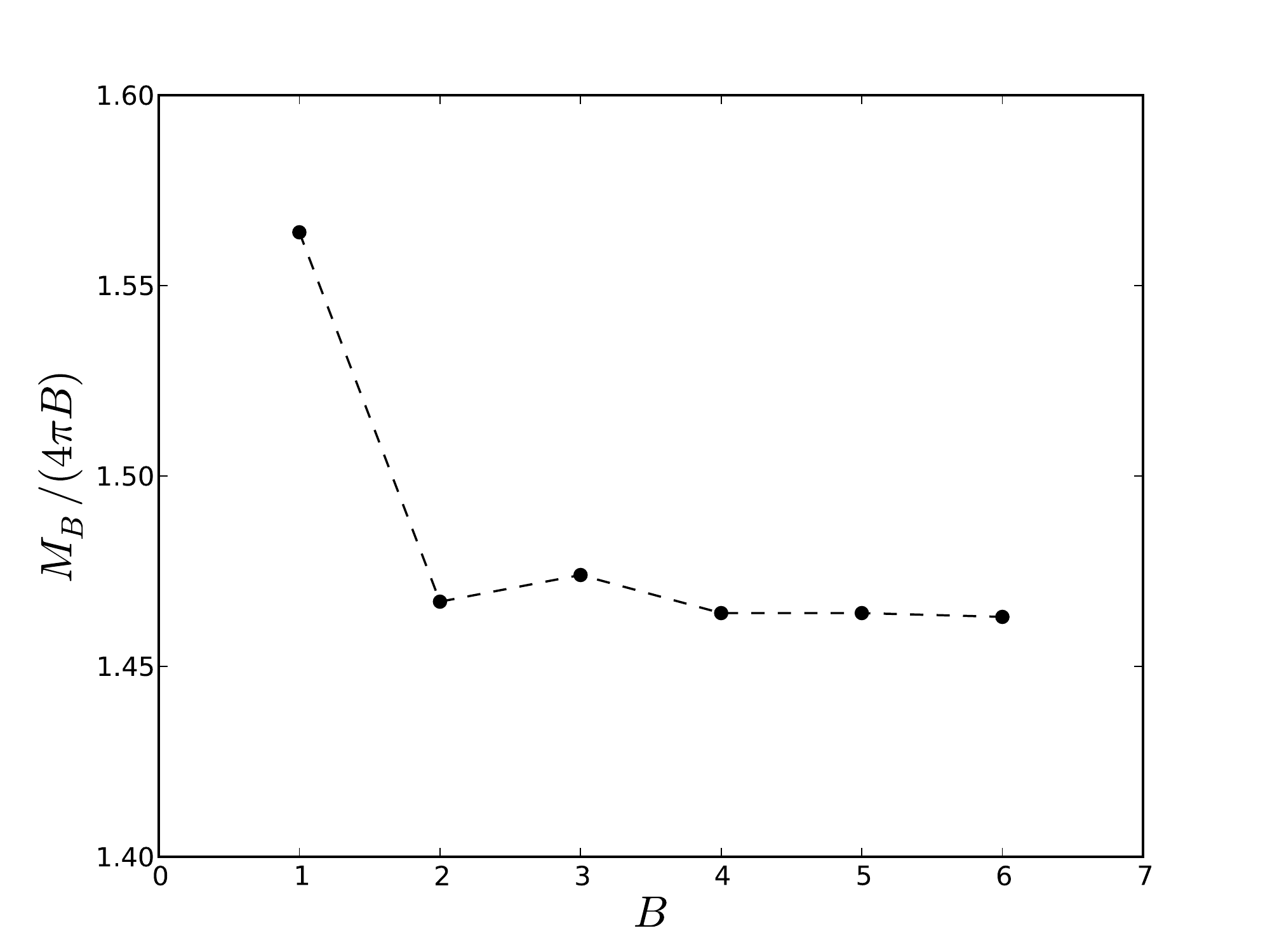}
\caption{Normalized energies $M_B/4\pi B$ vs baryon number $B$ for baby Skyrme solitons with potential (\ref{Lag_pot}) included. We plot the energy values of those configurations which we believe to be global energy minima.}
\label{Fig_Ebound_Baby_old}
\end{figure}

In Fig.~\ref{Fig_Ebound_Baby_old} we display the normalized energies per $4\pi B$ as function of the baryon number $B$ for the baby Skyrme configurations which we assume to be global energy minima.

\section{Numerical Results for Isospinning Baby Skyrmions}\label{Sec_Baby_Spin_Res}

In this section, we present the results of our 2D energy minimization simulations of isospinning soliton solutions in the baby Skyrme model with the potential choice (\ref{Lag_pot}). To find the stationary points of the total energy functional $H=M_B+K^2/(2\Lambda_B)$ for a fixed angular momentum $K$ and for a given topological charge $B$, we perform a 2D gradient flow evolution in analogy to Eq.~(\ref{Flow_eq}) starting with  well-chosen initial configurations. We use the static configurations obtained in the previous section as our start configurations for vanishing angular momentum ($K=0$). Then we increase $K$ in a stepwise manner using previously calculated configurations as starting configurations for the next value of $K$. To check our computations we also performed gradient flow calculations to minimize the pseudoenergy functional $F_\omega$ (\ref{Baby_Pseudo_energy_phi}) for a fixed rotation frequency $\omega$. Note that all simulation parameters are chosen as stated in Sec.~\ref{Sec_Baby_Static}. In particular, we carry out most of our simulations on relatively large grids containing $(601)^2$ lattice points with a lattice spacing $\Delta x=0.2$ in order to capture the asymptotic behaviour of our isospinning soliton solutions.

\begin{figure}[!htb]
\centering
\subfigure[\,Total energy vs angular frequency ]{\includegraphics[totalheight=5.cm]{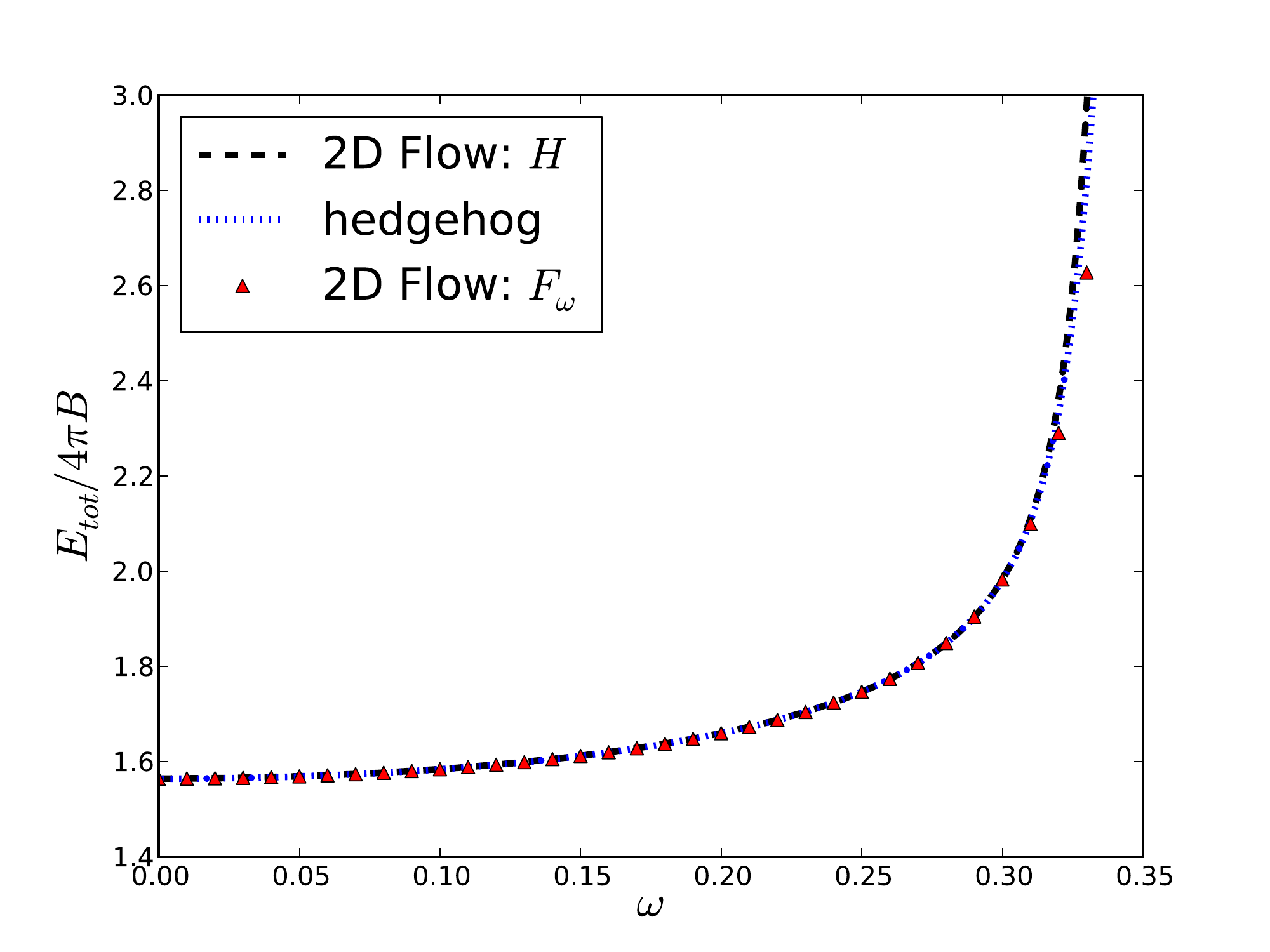}}
\subfigure[\,Mass-Isospin relationship ]{\includegraphics[totalheight=5.cm]{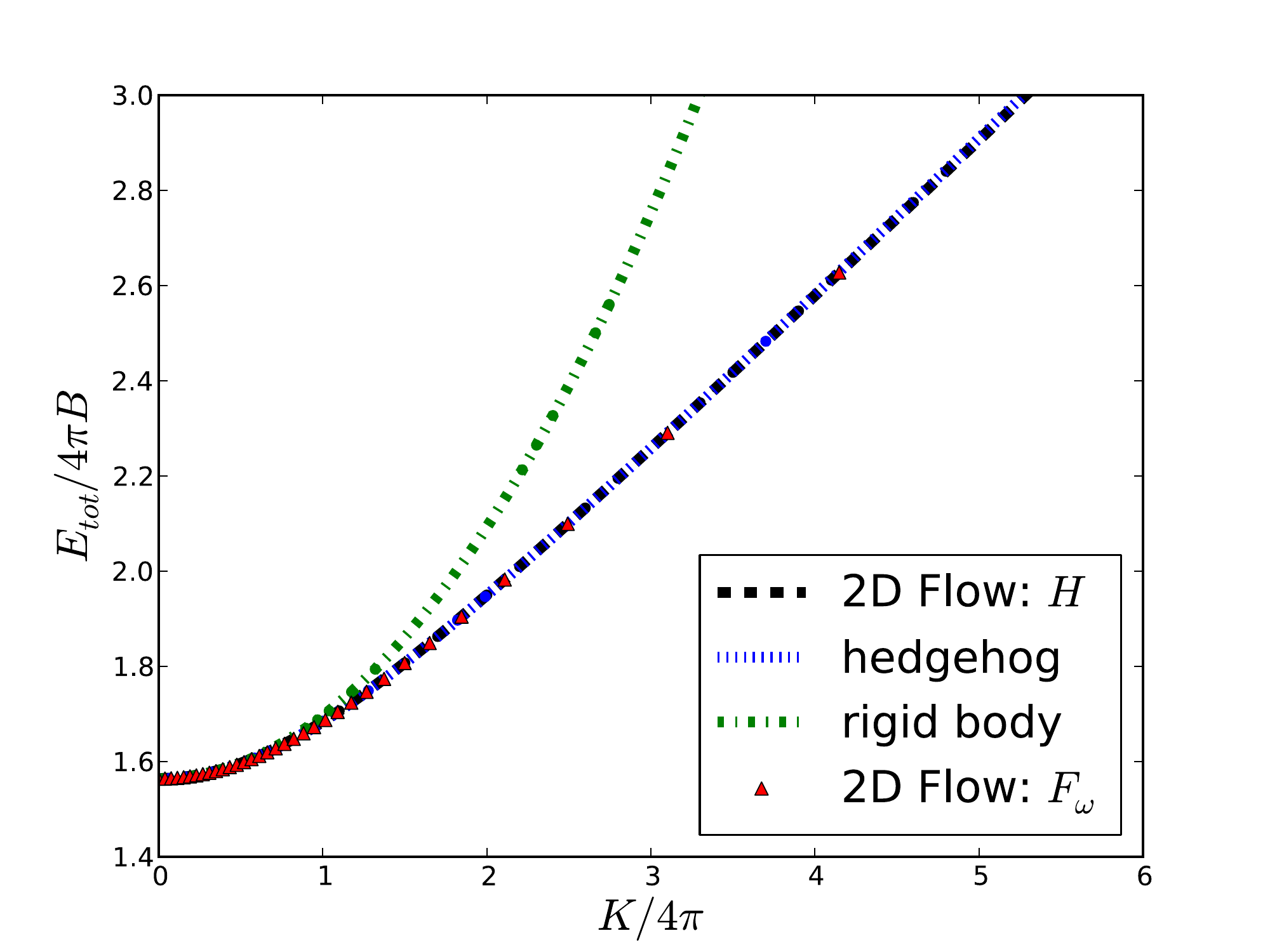}}\\
\subfigure[\,Inertia vs angular frequency]{\includegraphics[totalheight=5.cm]{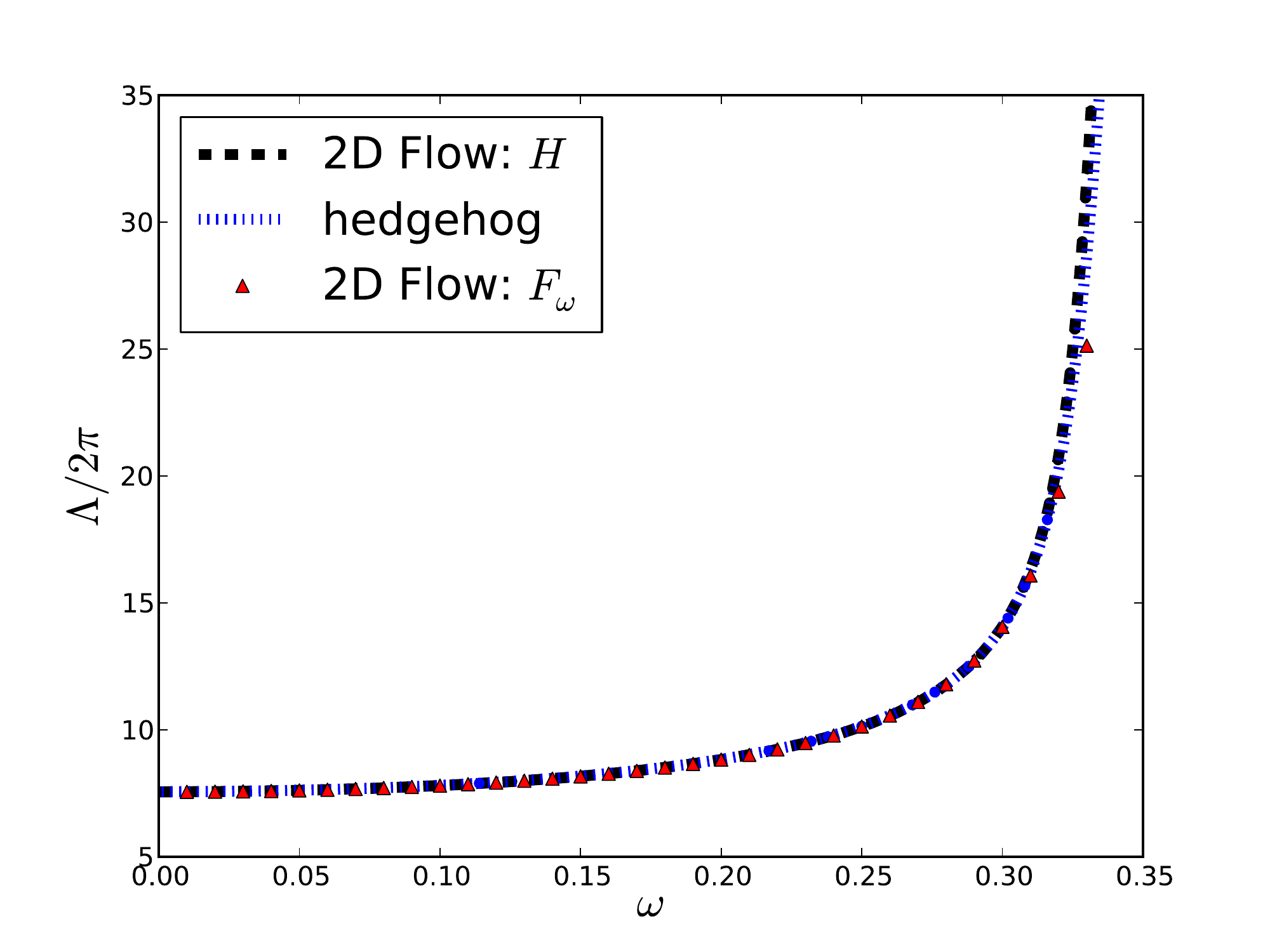}}
\subfigure[\,Inertia-Isospin relationship]{\includegraphics[totalheight=5.cm]{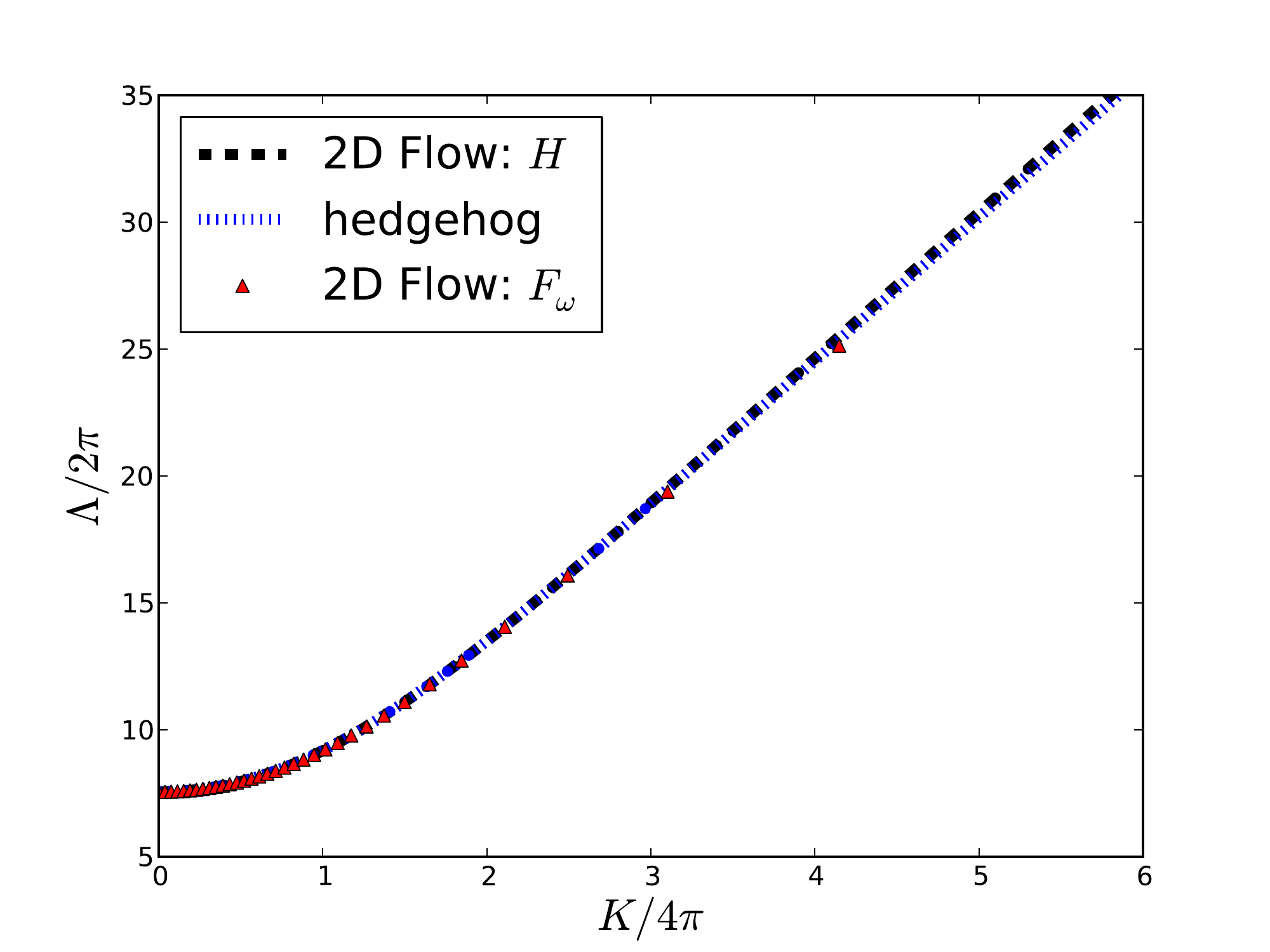}}
\caption{Isospinning $B=1$ soliton solution ($\mu=\sqrt{0.1}$). A starting configuration (\ref{Baby_hedge}) is numerically minimized using 2D gradient flow on a $(401)^2$ grid with a lattice spacing of $\Delta x=0.15$ and a time step size $\Delta t=0.005$. To check our numerics we explicitly verify that minimization of the pseudoenergy $F_\omega$ (for fixed angular frequency $\omega$) and minimization of the Hamiltonian $H$ (for fixed isospin $K$) reproduce the same curves. Additionally, we compare our results with those we expect for an isospinning, rotationally symmetric deforming 1-soliton solution. Our results agree well with those presented in \cite{Piette:1994mh}.}
\label{Ew_B1_V1}
\end{figure}

The majority of our computations of isospinning baby Skyrmions with charges $B=1-6$ have been performed with the rescaled mass parameter $\mu$ set to $\sqrt{0.1}$. In this case there is a maximum angular frequency $\omega_{\text{crit}}=\omega_2=\mu=\sqrt{0.1}$ beyond which no stable isospinning baby Skyrmion solution exists. 
\begin{figure}[!htb]
\centering
\subfigure[\,Total energy vs angular frequency]{\includegraphics[totalheight=5.cm]{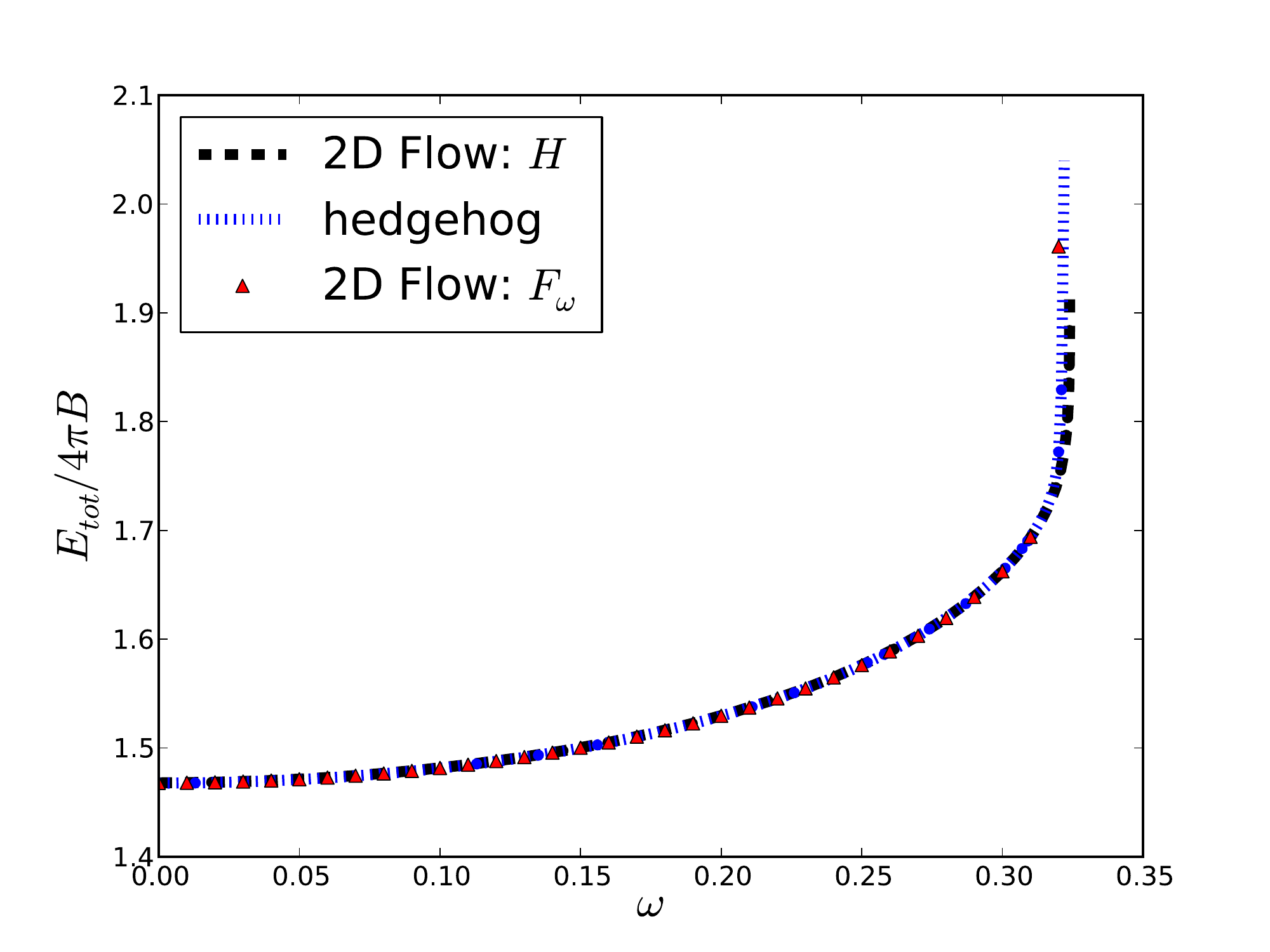}}
\subfigure[\,Mass-Isospin relationship]{\includegraphics[totalheight=5.cm]{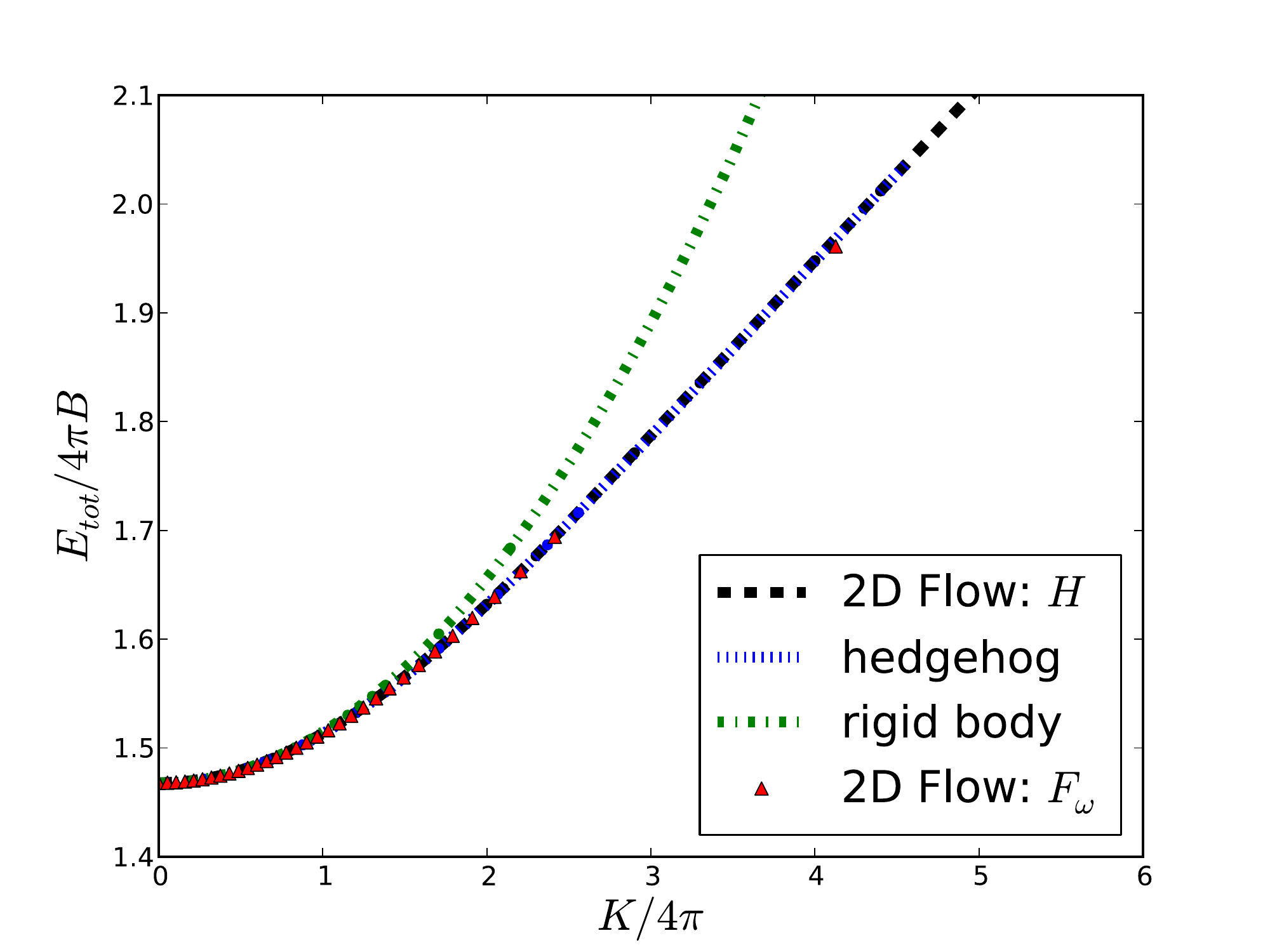}}\\
\subfigure[\,Inertia vs angular frequency]{\includegraphics[totalheight=5.cm]{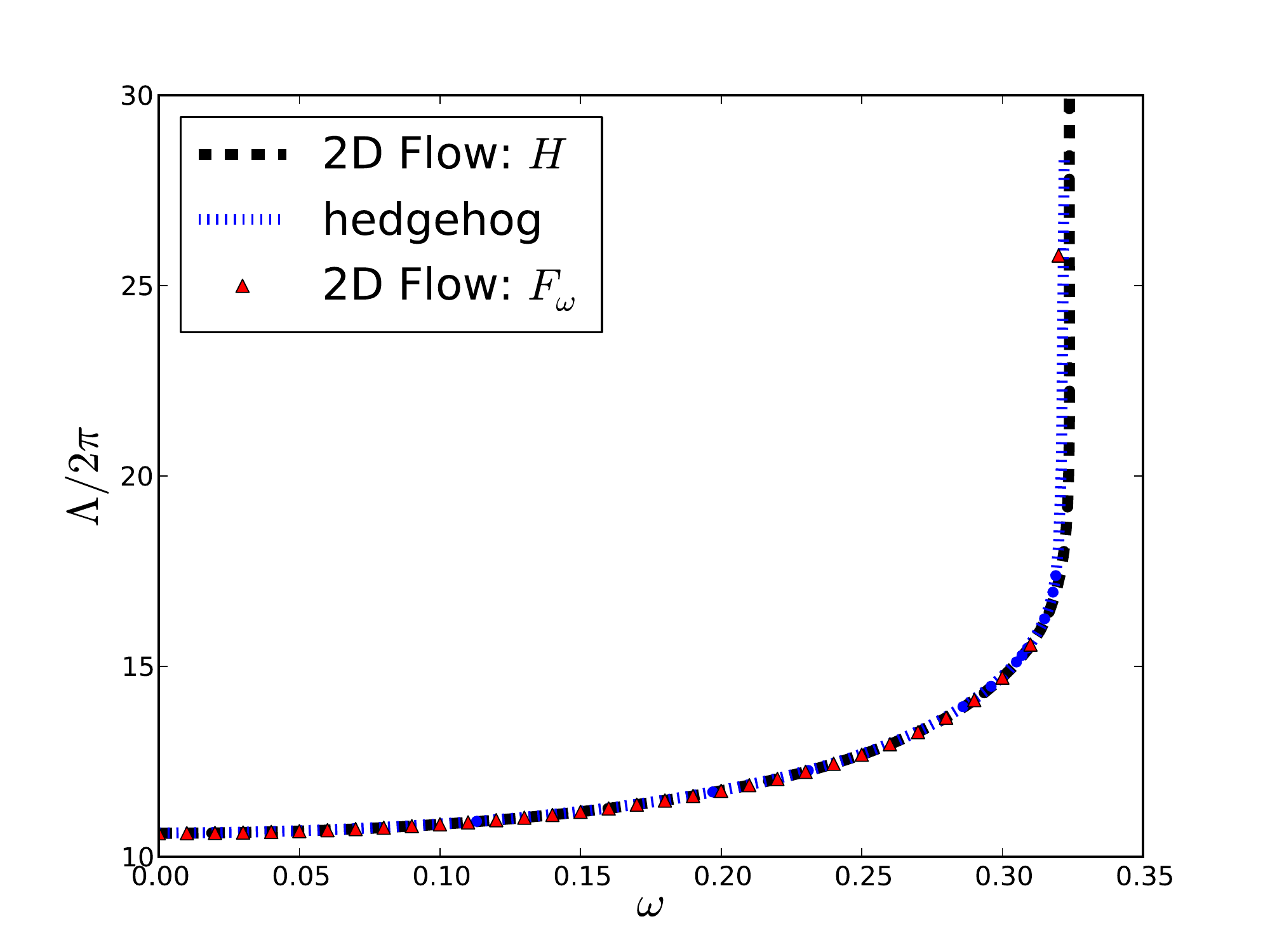}}
\subfigure[\,Inertia-Isospin relationship]{\includegraphics[totalheight=5.cm]{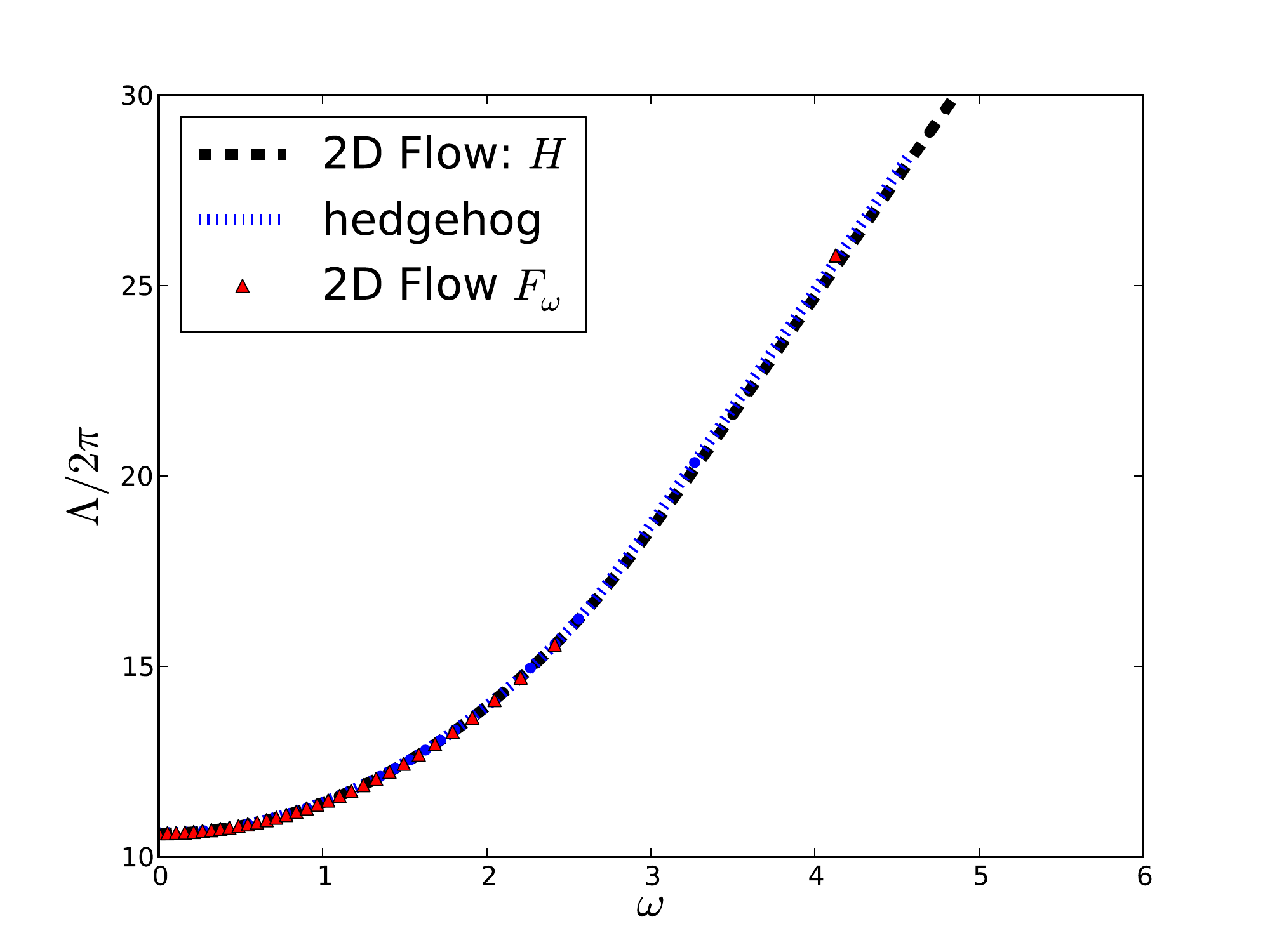}}
\caption{Isospinning $B=2$ soliton solution ($\mu=\sqrt{0.1}$). A starting configuration (\ref{Baby_hedge}) is numerically minimized using 2D gradient flow on a $(601)^2$ grid with a lattice spacing of $\Delta x=0.2$ and a time step size $\Delta t=0.005$. Our results agree perfectly with those obtained for an isospinning 2-soliton deforming within a hedgehog ansatz (\ref{Baby_hedge}).}
\label{Ew_B2_V1}
\end{figure}

\begin{itemize}
\item $B=1,2$: For the rotationally symmetric 1- and 2- soliton solutions we verify that our 2D gradient flow evolution reproduces, for the mass value $\mu=\sqrt{0.1}$, the behavior we expect from an isospinning hedgehog soliton solution \cite{Piette:1994mh}. Recall that for fields of the hedgehog type (\ref{Baby_hedge}) spatial rotations and isorotations are equivalent, and consequently $K$ can be interpreted as the total isospin or equivalently as the total spin of the baby Skyrme field $\boldsymbol{\phi}$. We plot in Figs.~\ref{Ew_B1_V1} and \ref{Ew_B2_V1} the dependencies of the spinning $B=1,2$ baby Skyrmion's mass, $E_{\text{tot}}$, and its moment of inertia, $\Lambda$, on its angular frequency, $\omega$, and its isospin, $K$. We confirm that we obtain the same energy and moment of inertia curves when we substitute the hedgehog field (\ref{Baby_hedge}) in the pseudoenergy functional $F_\omega$ or in the Hamiltonian $H$ and solve the associated variational equation \cite{Piette:1994mh}
\begin{align}
&\left(r+\left(\frac{B^2}{r}-\omega^2r\right)\sin^2f\right)f^{\prime\prime}+\left(1-\left(\omega^2+\frac{B^2}{r^2}\right)\sin^2f+\left(\frac{B^2}{r}-\omega^2r\right)f^\prime\sin f\cos f\right) f^\prime\no\\
&-\left(\frac{B^2}{r}-\omega^2r\right)\sin f \cos f-r\mu^2\sin f=0\,,
\label{ODE_hedgehog}
\end{align}
for the radial profile function $f(r)$ with boundary conditions $f(0)=\pi$ and $f(\infty)=0$. 

$E_{\text{tot}}(\omega)$ and $\Lambda(\omega)$ grow rapidly with $\omega$ and diverge at $\omega_{\text{crit}}\approx\sqrt{0.1}$. We can see that the total energy $E_{\text{tot}}(K)$ and the moment of inertia $\Lambda(K)$ increase linearly with $K$ for $K$ sufficiently large and only depend quadratically on $K$ for slowly rotating solutions. Consequently, the rigid-body approximation is only a good approximation for small values of $K$, in particular, $K\leq1.05\times4\pi$ for $B=1$ and $K\leq1.51\times4\pi$ for $B=2$. Close to the cutoff ($\omega\approx0.31$, $K\approx2.49\times4\pi$) the energy values given by the rigid-body formula are roughly $10\%$ larger than those for the non-rigidly rotating 1-soliton solution. Similarly, for $B=2$ the rigid-body approximation predicts an energy value at $\omega\approx0.31,\,K\approx2.49\times4\pi$ which is approximately $3\%$ larger than the one calculated for the deforming charge-2 solution. The  energy density contour plots  in Fig.~\ref{Fig_en_dens_V1_1} show the deformation of the charge-1 and charge-2 solitons as function of isospin $K$ and rotation frequency $\omega$. As shown in Fig.~\ref{Fig_en_dens_V1_1}, the isospinning charge-2 configuration preserves its rotational symmetry for all frequency values  $\omega<0.31$ and spontaneously breaks its rotational symmetry at $\omega\approx \omega_2$.

\begin{figure}[!htb]
\centering
\includegraphics[totalheight=11.0cm]{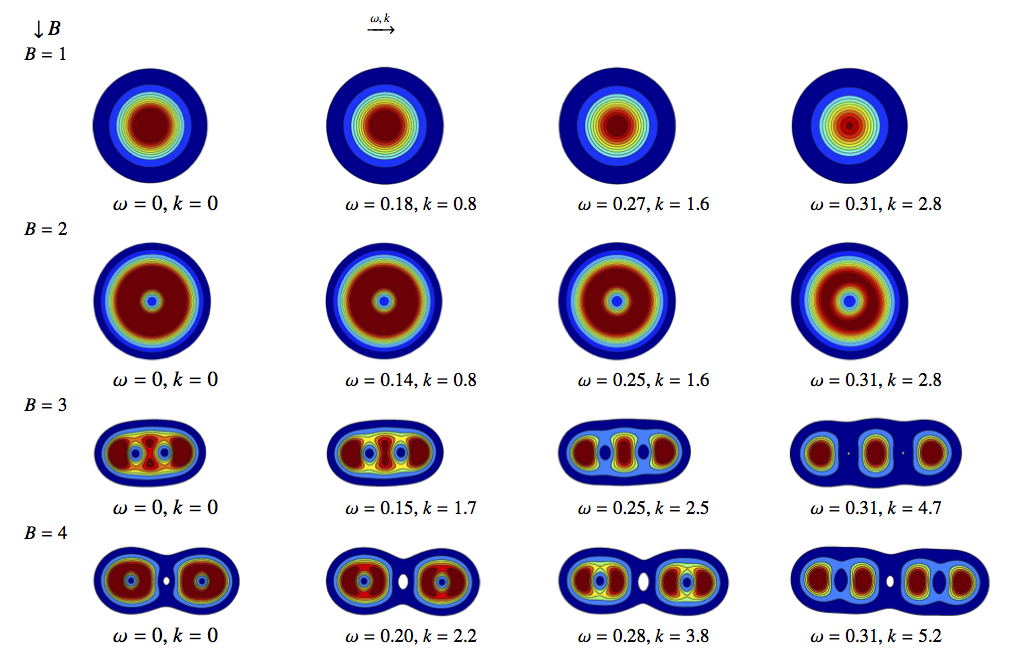}
\caption{Energy density contour plots for isospinning multisolitons in the standard baby Skyrme model  with charges $B=1-4$ and mass parameter $\mu$ chosen to be $\sqrt{0.1}$. To simplify comparison with \cite{Piette:1994mh}, the isospin $K$ is given in units of $4\pi$, i.e. we define $k=K/4\pi$.}
\label{Fig_en_dens_V1_1}
\end{figure}

In the  $(3+1)$-dimensional Skyrme-Faddeev model it was found \cite{Harland:2013uk} that stable, isospinning soliton solutions only exist for angular frequencies $\omega\leq \text{min}\left\{1,\mu\right\}$ and that they can be destabilized by nonlinear velocity terms in the field equations far before the upper limit $\omega_2=\mu$ is reached. Similarly, we find in the baby Skyrme model that for $\mu>1$ isospinning soliton solutions become unstable far before reaching $\omega_2$. Fully two-dimensional relaxation calculations \emph{reveal} that stable, isospinning soliton solutions only exist for angular frequencies $\omega\le\text{min}\left(\mu,1\right)$. We display in Fig.~\ref{Fig_B1_Old_mass} the critical behaviour of isospinning charge-1 Skyrmion solutions for a range of mass values $\mu$. 

For the mass range $\mu\in(0,1]$ the total energy, $E_{\text{tot}}$, and the moment of inertia, $\Lambda$, \emph{diverge} at $\omega_2=\mu$, whereas for mass values $\mu>1$ we find that they take \emph{finite} values at the critical frequency $\omega_1=1$. It is interesting to compare this pattern of critical behaviour with the one calculated when only considering rotationally symmetric deformations. As shown in the Appendix, the so-obtained critical frequencies for charge-1 and -2 solitons inadvertently suggest the existence of stable isospinning baby Skyrmion solutions with mass $\mu>1$ for angular frequencies $\omega>\omega_1=1$. This result is simply an artefact of the hedgehog approximation (\ref{Baby_hedge}).

\begin{figure}[!htb]
\centering
\begin{tabularx}{\textwidth}{XXX}
\epsfig{file=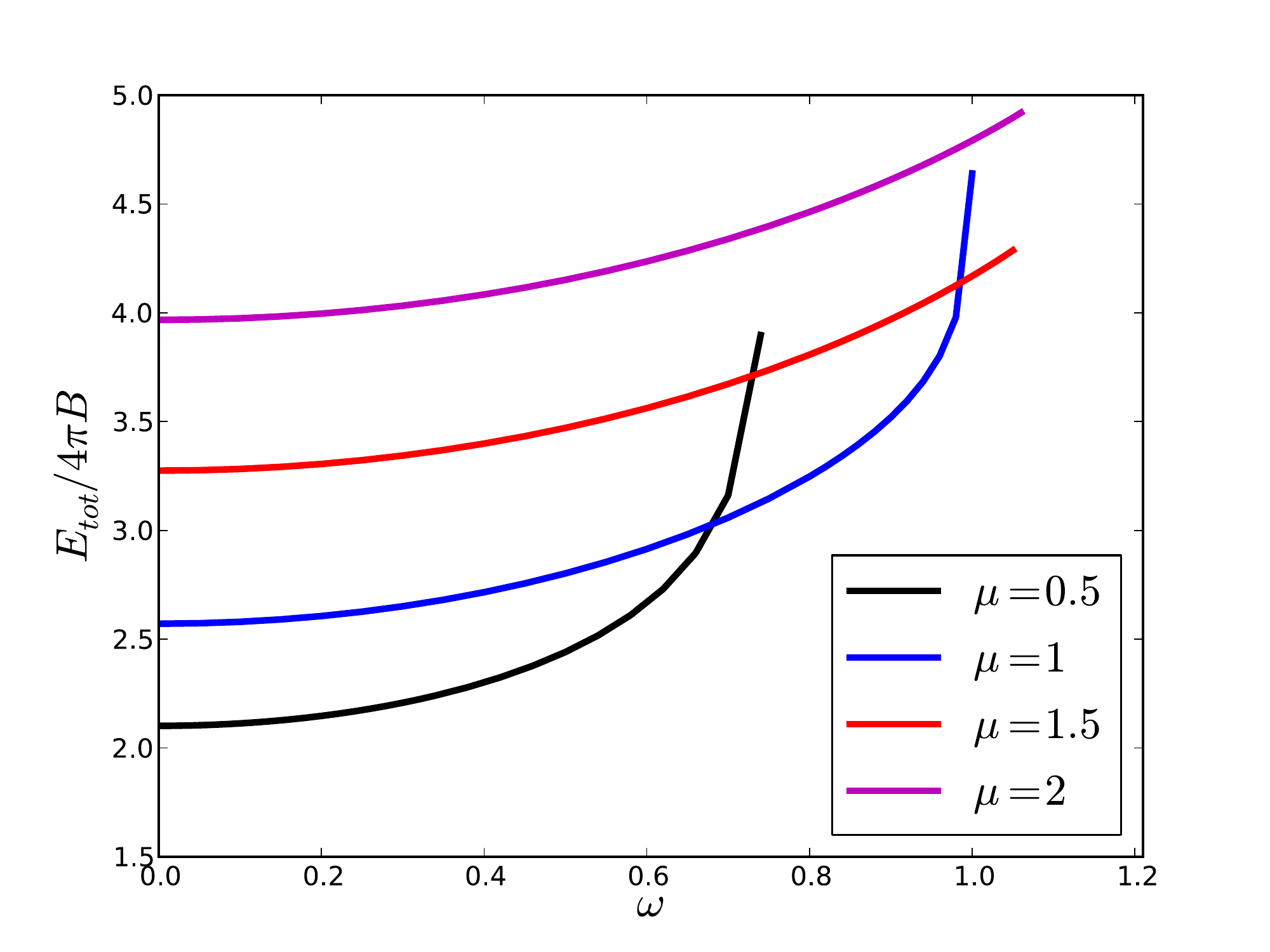,totalheight=4.0cm,clip=} &
\epsfig{file=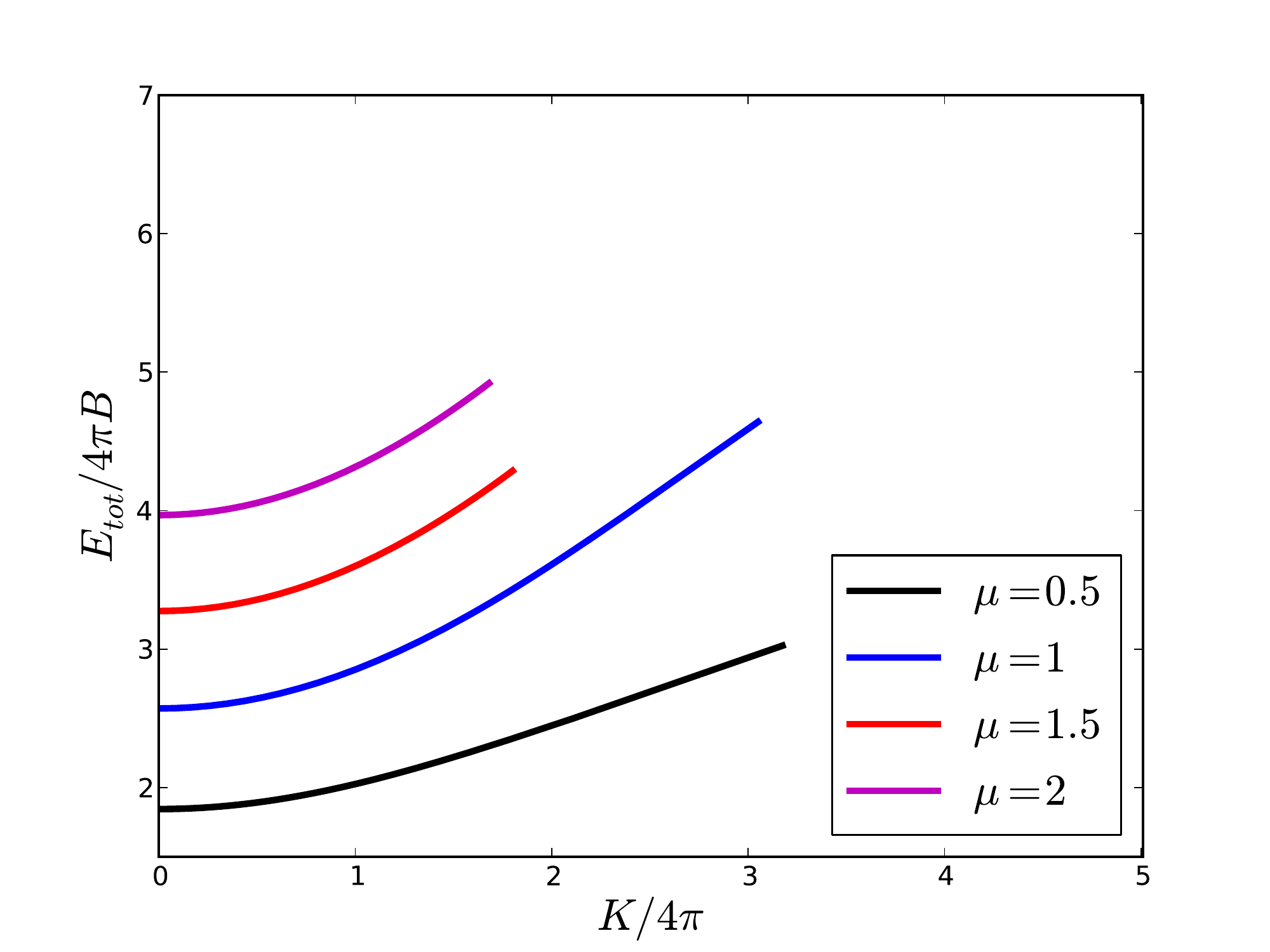,totalheight=4.0cm,clip=} &
\epsfig{file=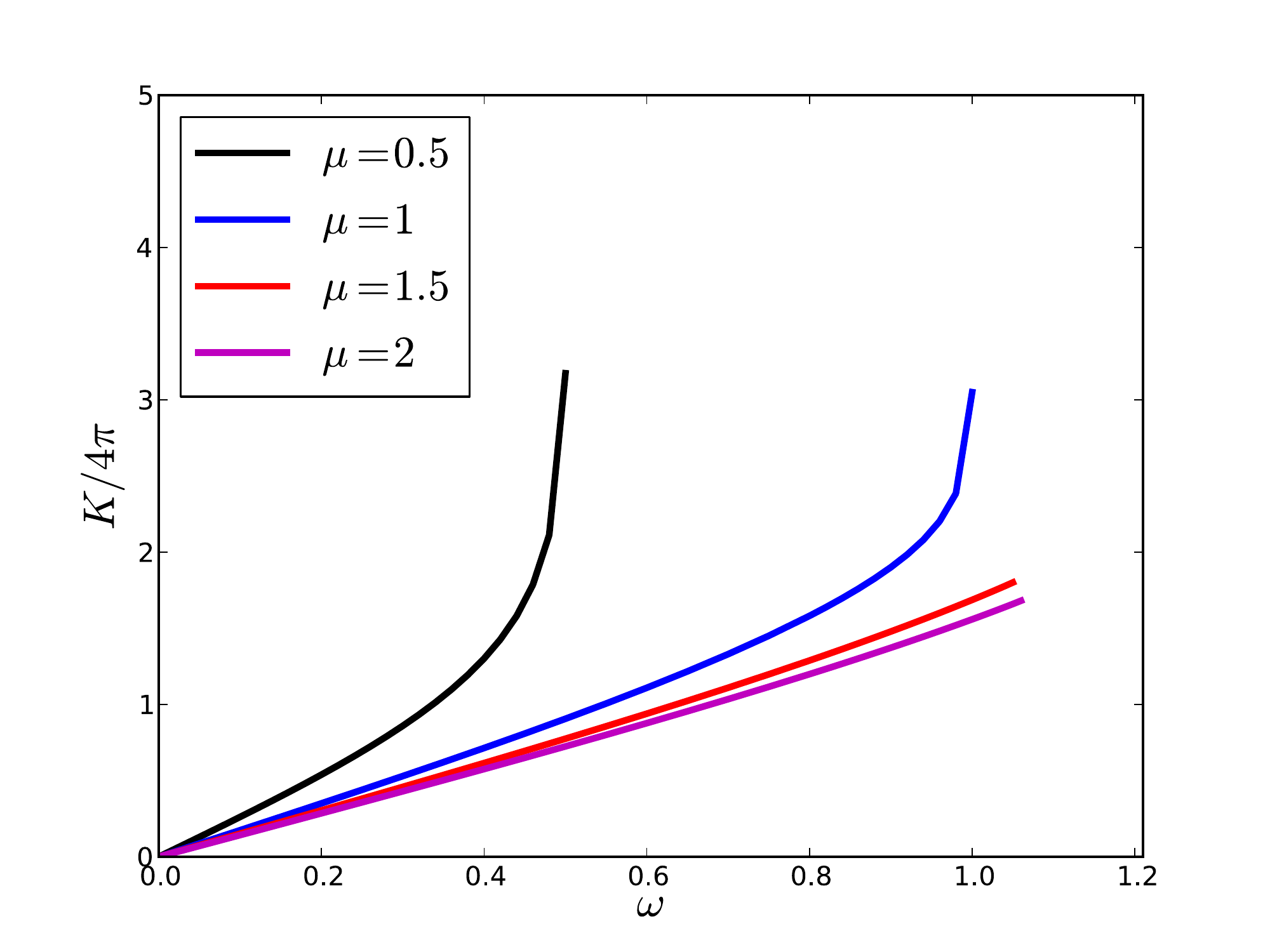,totalheight=4.0cm,clip=}
\end{tabularx}
\caption{Total energy $E_{\text{tot}}$ and isospin $K$ for $B=1$ soliton solutions in the standard baby Skyrme model as a function of angular frequency $\omega$. The mass parameter takes the values $\mu=0.5,1,1.5,2$. The same pattern of critical behaviour can be confirmed numerically for isospinning multisoliton solutions.}
\label{Fig_B1_Old_mass}
\end{figure}

Furthermore, for larger mass values ($\mu>\sqrt{0.1}$) our full two-dimensional relaxation calculations show that at some third critical angular frequency value, $\omega_3$, the isospinning charge-2 soliton solutions become unstable to breaking up into their charge-1 components which start moving  apart from each other. Generally speaking, increasing the mass value $\mu$ results in increasingly larger rotational symmetry breaking at a given angular momentum $K$ (see Fig.~\ref{Fig_B1B2_hedge_com}). Rotationally symmetric Skyrme configurations are found to be of significantly higher energy and turn out to be unstable for sufficiently large $\mu$ and $K$.  The corresponding energy density contour plots for a range of values of $\mu$ are shown in Fig.~\ref{Fig_iso_B2_mass}. We observe that for increasing mass parameter $\mu$, the breakup into individual charge-1 constituents occurs at increasingly higher values of $\omega_3\le \omega_1=1$ (compare the middle plot in Fig.~\ref{Fig_B1B2_hedge_com} and the corresponding breakup frequency values listed in the table). Recall that isospinning baby Skymion solutions \emph{do not} minimize  the total energy functional $E_{\text{tot}}(\omega)$ for fixed angular frequency $\omega$, which explains the negative deviations $\Delta E_{\text{tot}}$ from the rotationally symmetric deforming charge-2 baby Skyrmion configuration shown in the middle plot of  Fig.~\ref{Fig_B1B2_hedge_com}.

\begin{figure}[!htb]
\centering
\begin{tabularx}{\textwidth}{XXX}
\epsfig{file=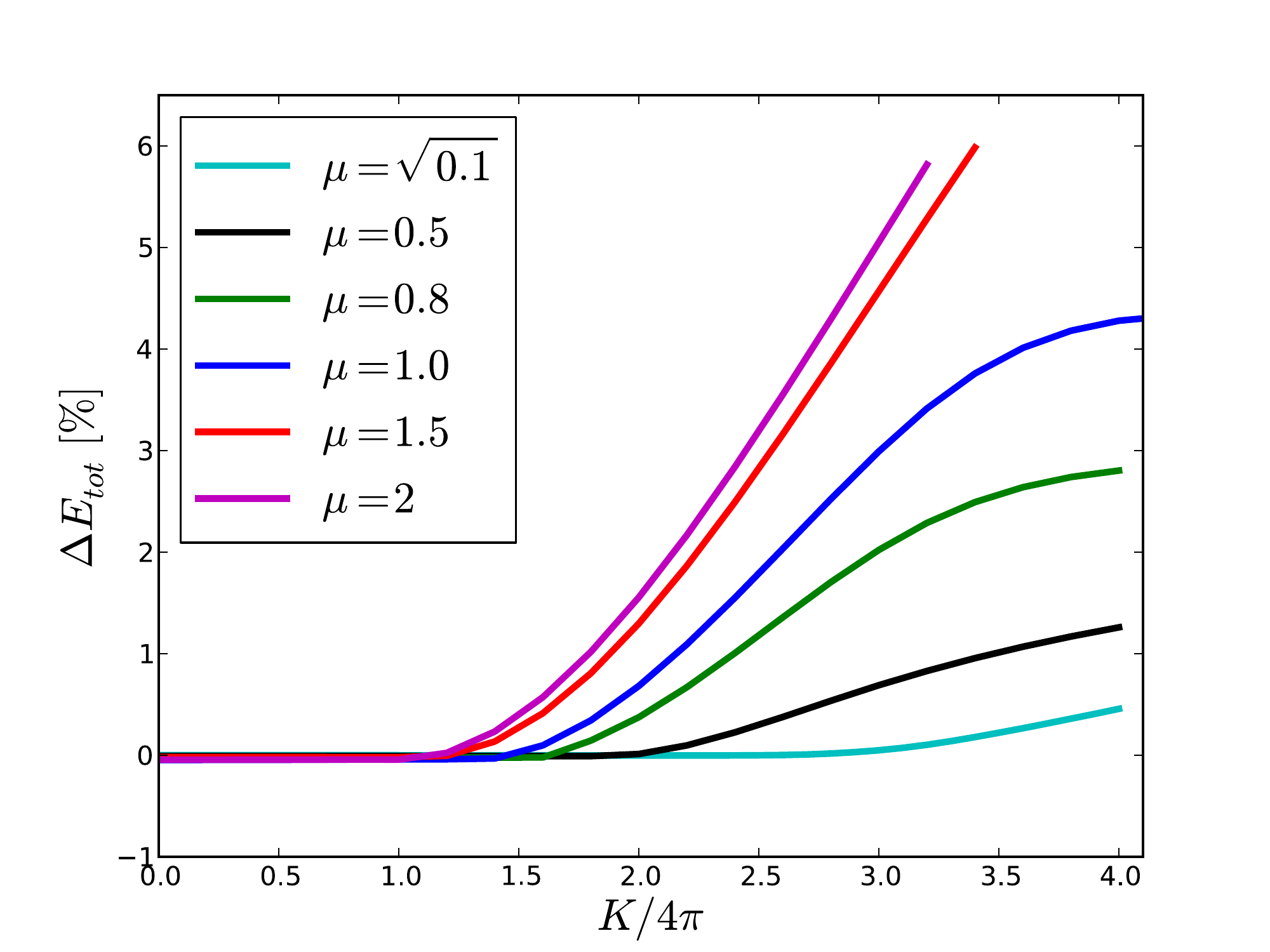,totalheight=4.5cm,clip=} &
\epsfig{file=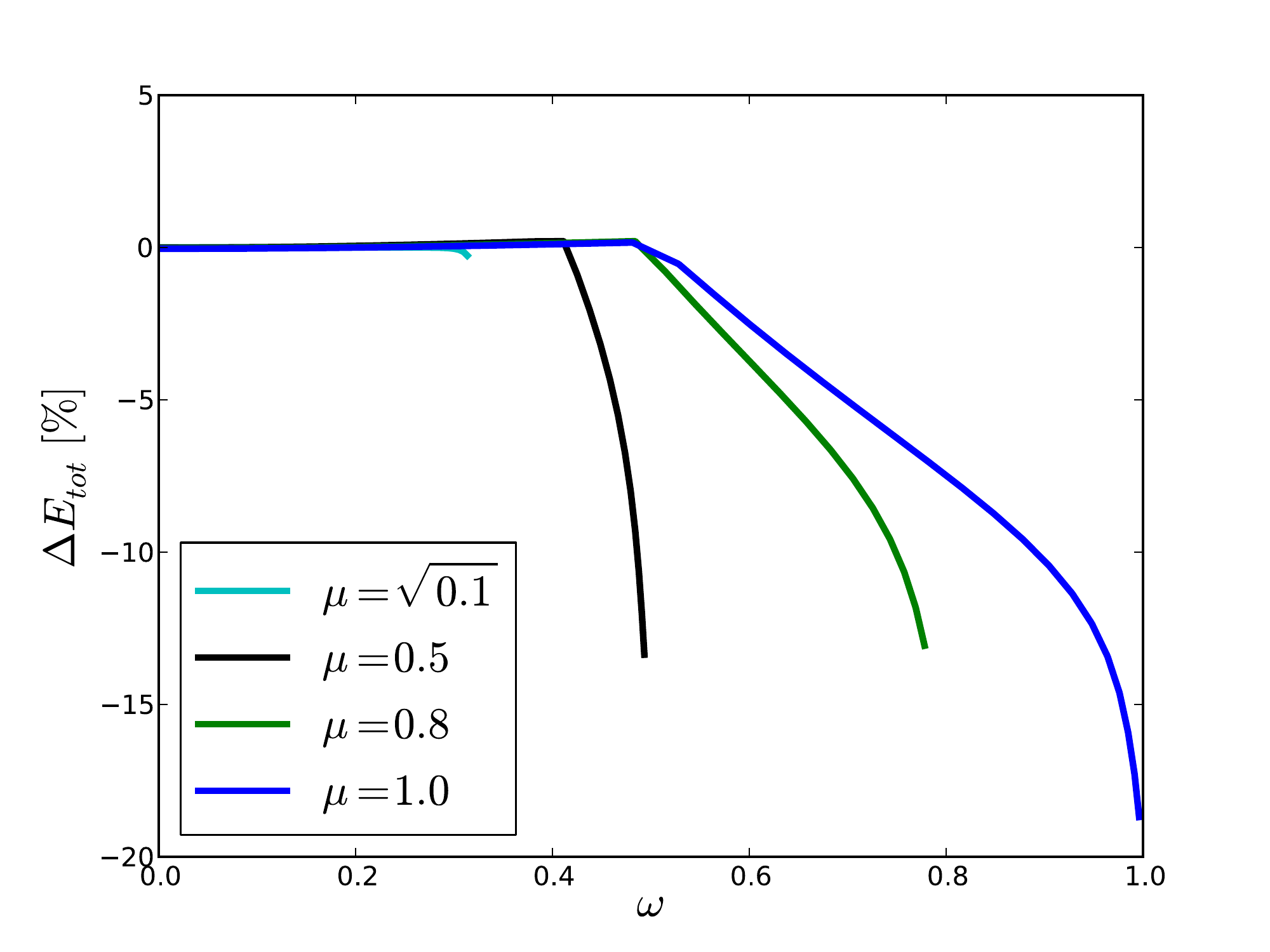,totalheight=4.5cm,clip=} &\vspace{-3cm}\hspace{0.1cm}
\begin{tabular}{|c|c|c|c|c|c|l|}\hline                                                                                                                                                                         $\mu$& $\sqrt{0.1}$&$0.5$ &$0.8$ &$1.0$ & $1.5$ &$2.0$\\\hline                                                                                                                                         
    $\omega_3$& $0.31$ &$0.41$&$0.48$ &$0.49$ & $0.50$ & $0.53$\\\hline
\end{tabular}
\end{tabularx}
\caption{The deviation $\Delta E_{\text{tot}}=\left(E_\text{hedgehog}-E_{\text{tot}}\right)/E_\text{tot}$  from the rotationally symmetric deforming charge-2 baby Skyrmion configuration (\ref{ODE_hedgehog}) as a function of isospin $K$ (left panel) and angular frequency $\omega$ (middle panel) for a range of mass values $\mu$. Approximate values for the break-up frequencies $\omega_3$ -- the frequencies at which the charge-2 solutions start to split into its charge-1 constituents -- are listed in the table on the right-hand side. We verified that for this mass range the isospinning charge-1 baby Skyrmion solution 
does not deviate significantly from a rotationally symmetric deforming $B=1$ Skyrme configuration. }
\label{Fig_B1B2_hedge_com}
\end{figure}

\begin{figure}[!htb]
\centering
\includegraphics[totalheight=18.0cm]{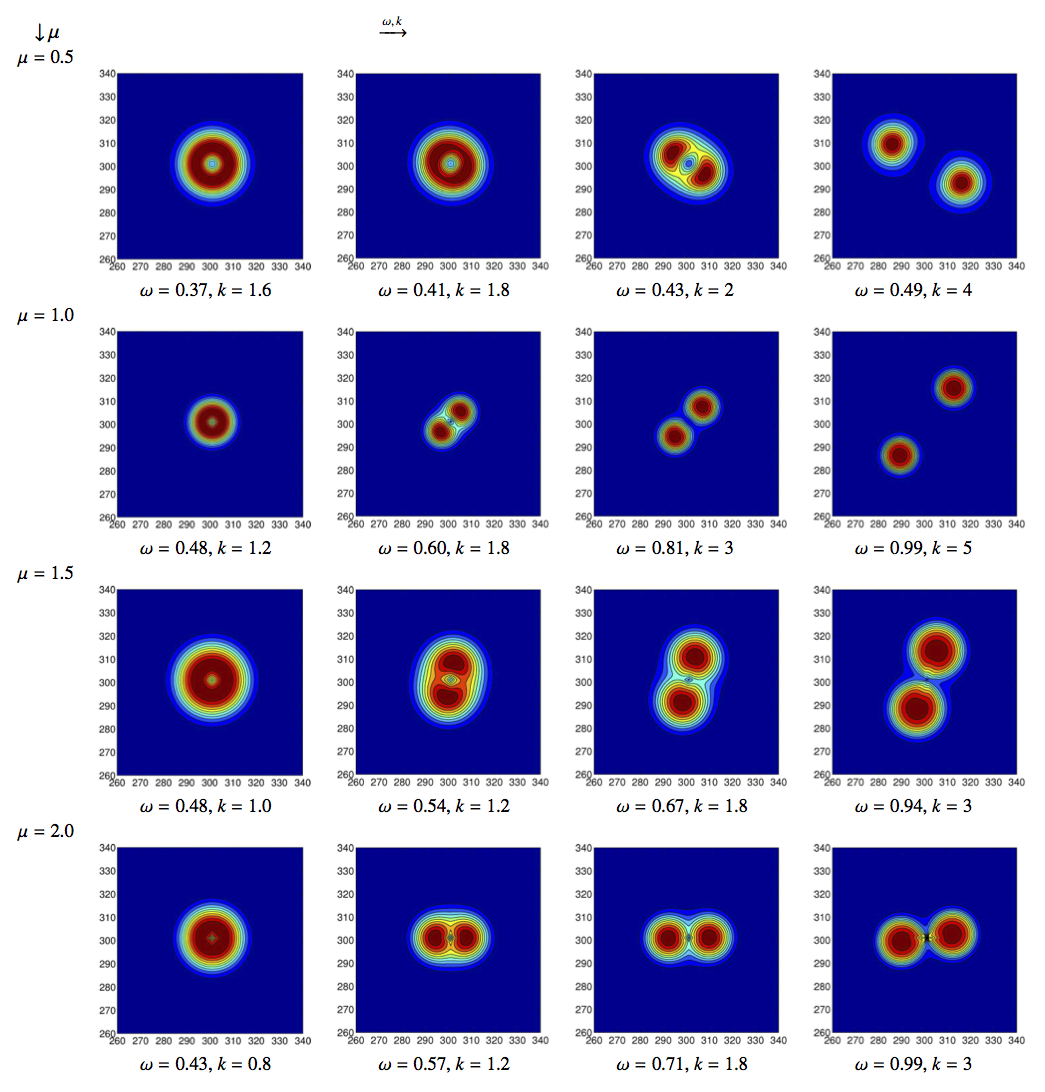}
\caption{Energy density contour plots for isospinning $B=2$ soliton solutions in the standard baby Skyrme model for a range of  mass values $\mu$. Note that the results presented here have been obtained using $(601)^2$ grids with lattice spacing $\Delta x=0.2$ for $\mu=0.5,1$ and $\Delta x=0.1$ for $\mu=1.5,2$.}
\label{Fig_iso_B2_mass}
\end{figure}

We show in Fig.~\ref{Fig_B1B2_Rigid_com} the deviations from the rigid body, plotted against the angular momentum $K$, for both the charge-1 and charge-2 solutions. As the mass value $\mu$ increases, the rigid-body approximation provides more accurate results for the isospinning solutions of the model.

\begin{figure}[!htb]
\centering
\subfigure[\,$B=1$]{\includegraphics[totalheight=6.cm]{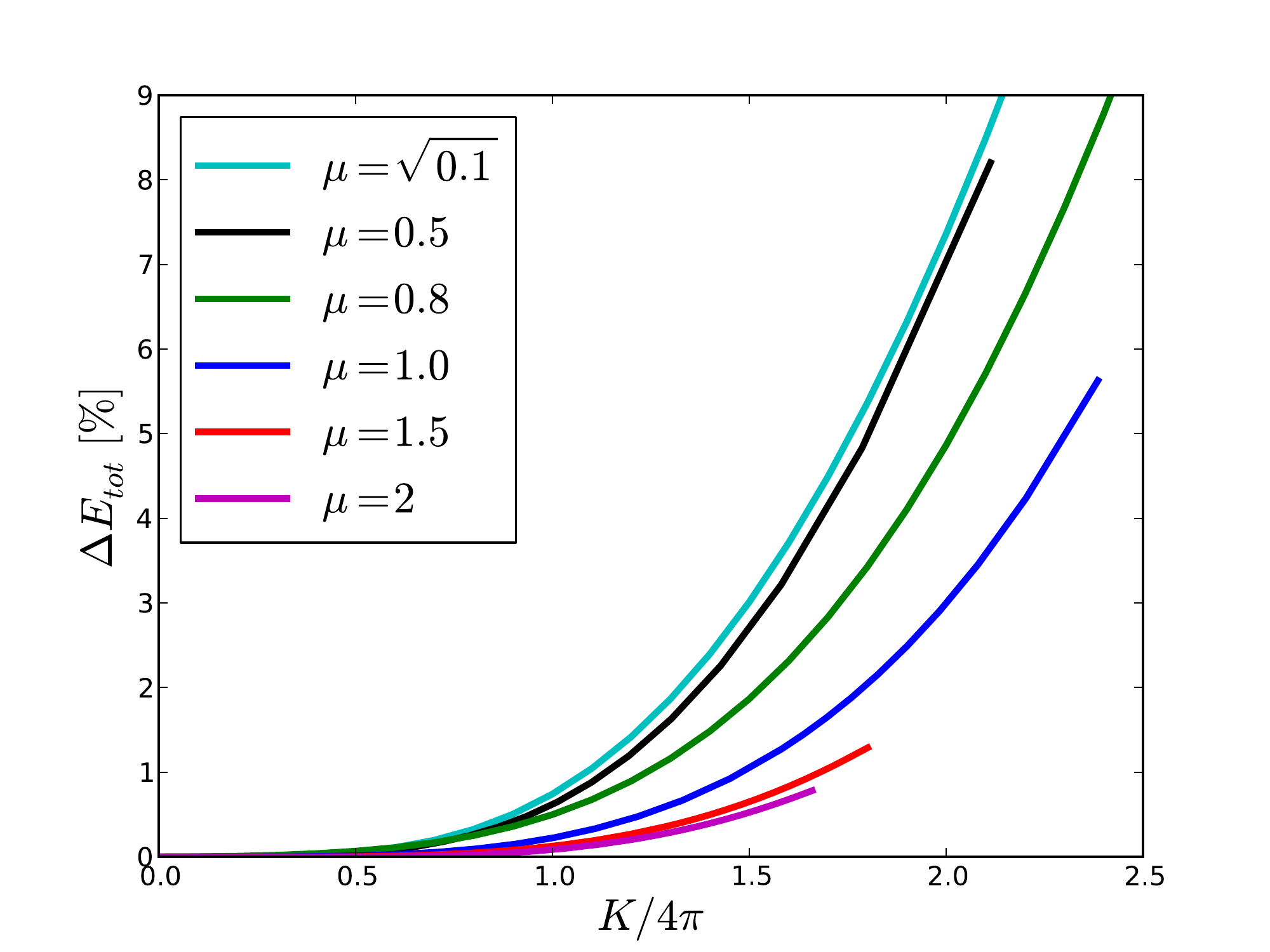}}
\subfigure[\,$B=2$]{\includegraphics[totalheight=6.cm]{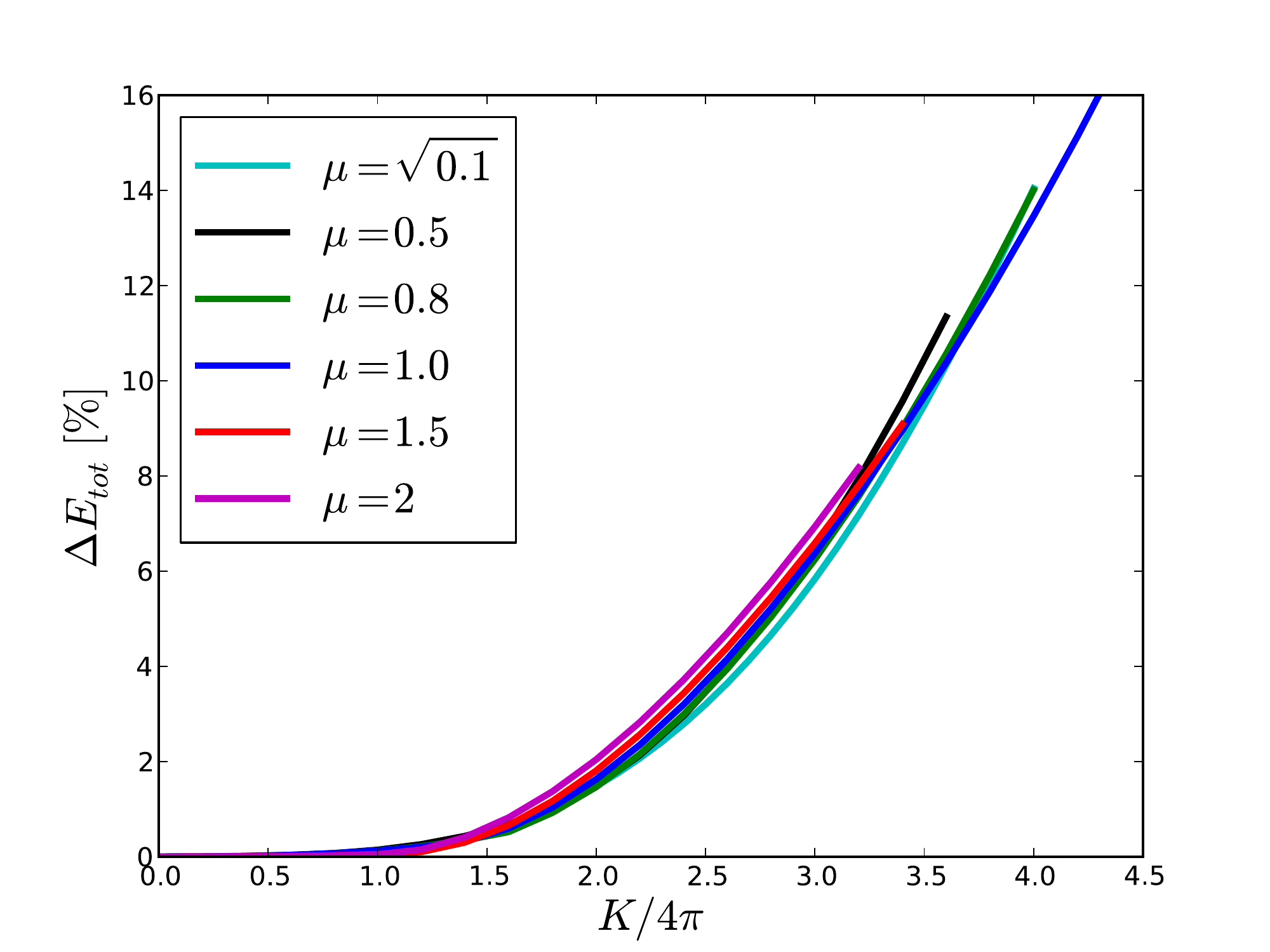}}
\caption{The deviation $\Delta E_{\text{tot}}=\left(E_\text{Rigid}-E_{\text{tot}}\right)/E_\text{tot}$ from the rigid-body approximation for charge-1 and charge-2 baby Skyrmions as a function of isospin $K$ for various rescaled mass values $\mu$.}
\label{Fig_B1B2_Rigid_com}
\end{figure}

\item $B=3$: For the standard mass value ($\mu=\sqrt{0.1}$) the linear 3-soliton splits into three weakly bound, linearly arranged  $B=1$ hedgehog solitons when isospinning about the $z$ axis, as shown by the energy density contour plots in Fig.~\ref{Fig_en_dens_V1_1}. The energy curves given in Fig.~\ref{EwJ_B3B4_V1} show the linear dependence of the total energy $E_{\text{tot}}(K)$ on the isospin $K$ for $K>2.0\times4\pi$. For $K<2.0\times4\pi$ deformations due to centrifugal effects can be neglected and the isospinning solution can be essentially seen as a rigid rotor. We find that with increasing mass value $\mu$ the isospinning $B=3$ soliton becomes increasingly stable to breaking up into its constituents; i.e. the break-up frequency $\omega_3$ takes larger values.

\begin{figure}[!htb]
\centering
\begin{tabularx}{\textwidth}{XXX}
\epsfig{file=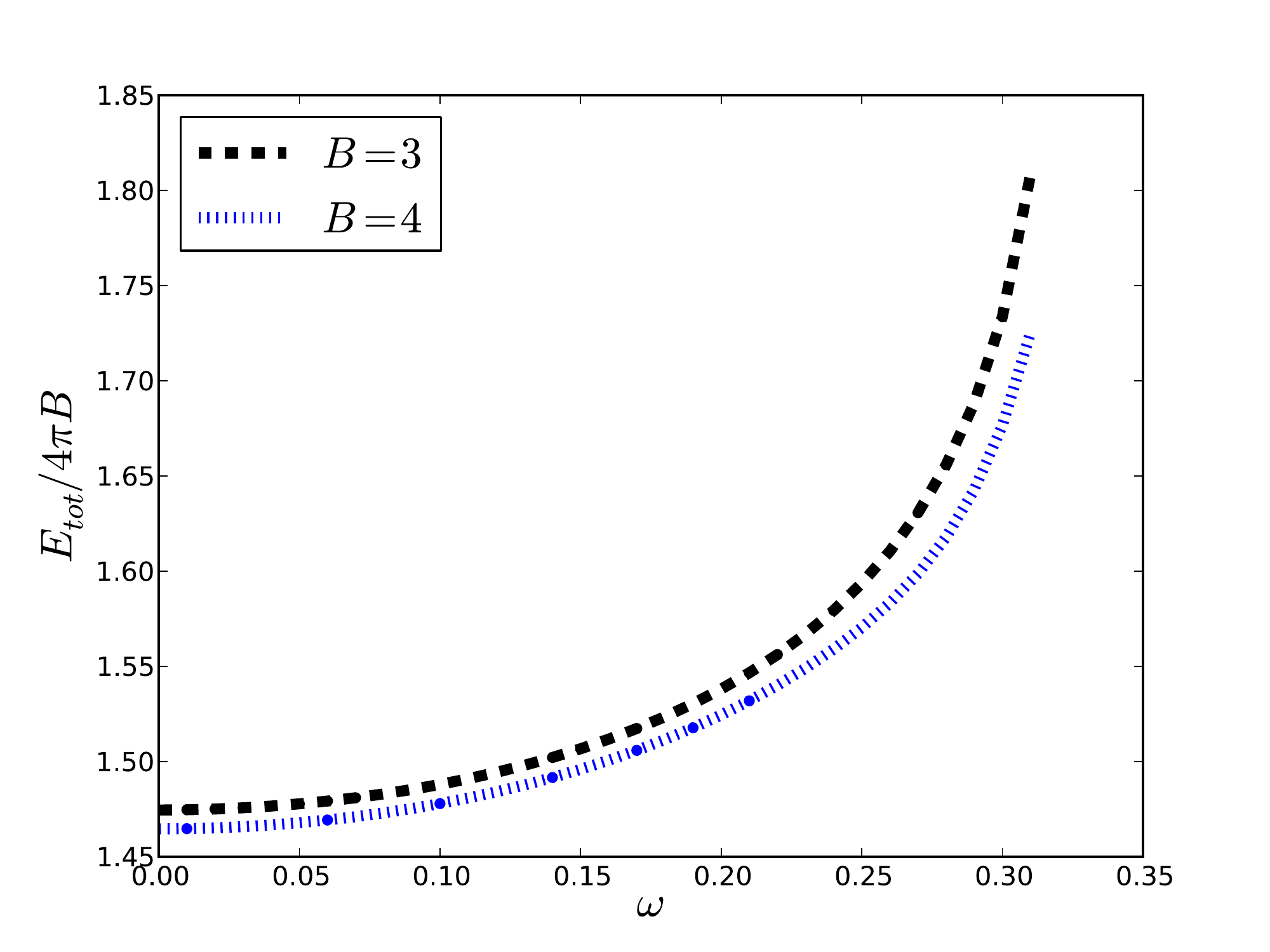,totalheight=4.0cm,clip=} &
\epsfig{file=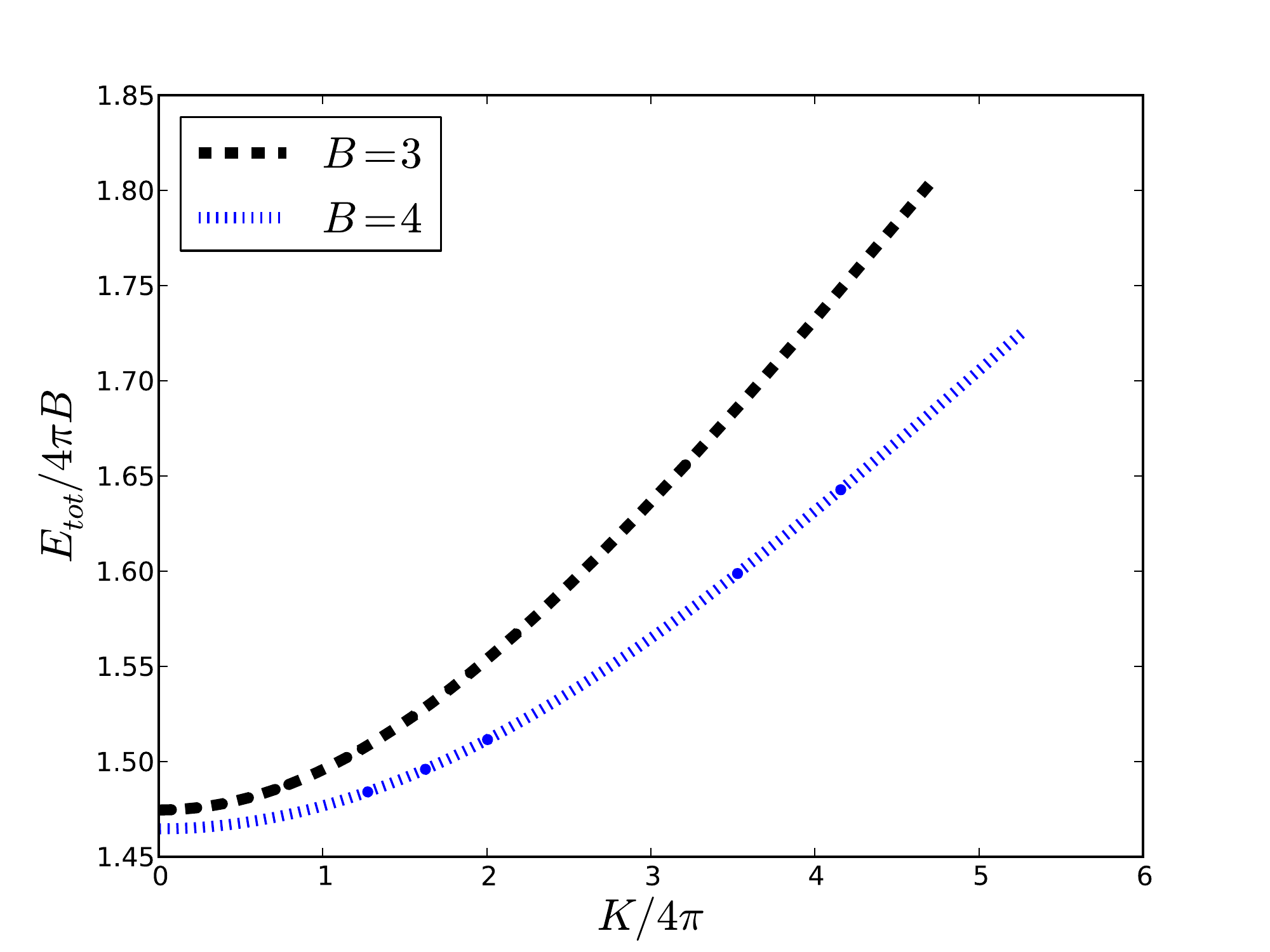,totalheight=4.0cm,clip=} &
\epsfig{file=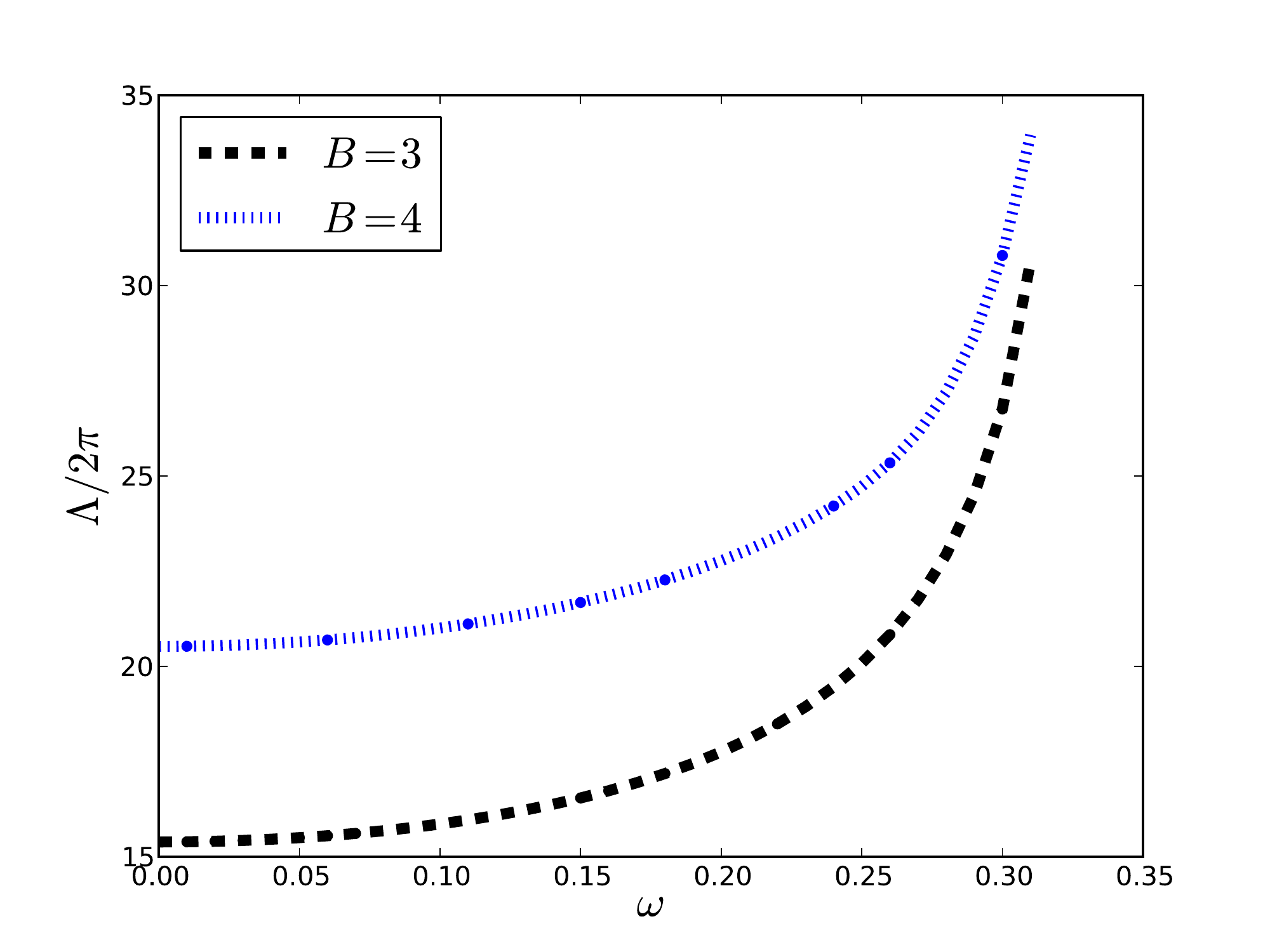,totalheight=4.0cm,clip=}
\end{tabularx}
\caption{Total energy $E_\text{tot}$ as a function of angular frequency $\omega$ and as a function of isospin $K$ for solitons ($\mu=\sqrt{0.1}$) in the conventional baby Skyrme model with baryon number $B=3,4$.}
\label{EwJ_B3B4_V1}
\end{figure}

\item $B=4$: The energy densities for the isospinning 4-baby Skyrme soliton are plotted for $\mu=\sqrt{0.1}$ in the last row of Fig.~\ref{Fig_en_dens_V1_1}. The pair of two weakly bound 2-solitons breaks into 4 single linearly arranged 1-solitons. The corresponding moment of inertia curves $\Lambda(\omega)$ and energy curves $E_{\text{tot}}(\omega),\,E_{\text{tot}}(K)$ can be found in Fig.~\ref{EwJ_B3B4_V1}. As $\mu$ increases the splitting into individual charge-1 constituents happens at increasingly higher rotational frequency values.

\begin{figure}[!htb]
\centering
\includegraphics[totalheight=12.0cm]{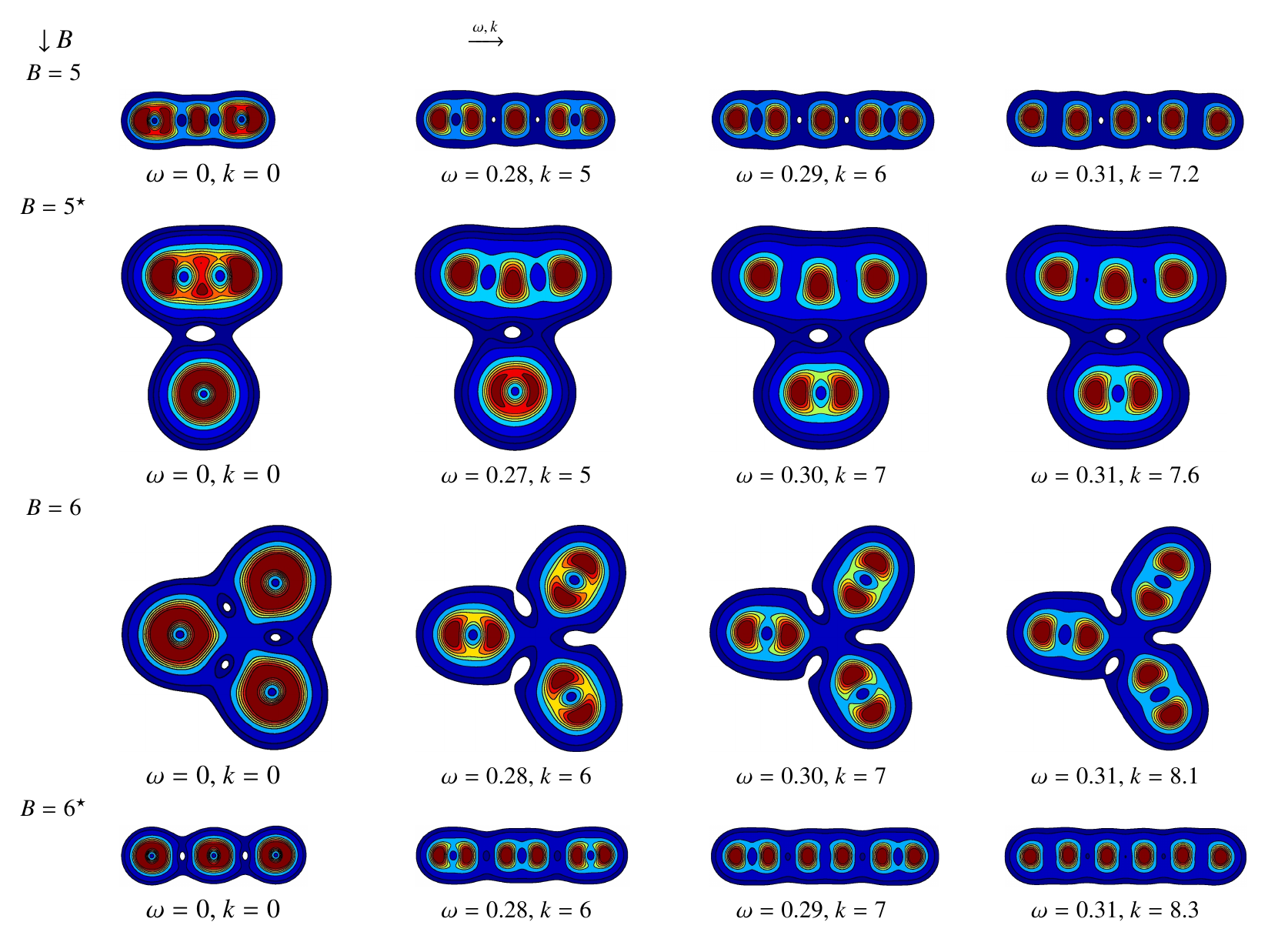}
\caption{Energy density contour plots for isospinning multisolitons in the standard baby Skyrme model with charges $B=5,6$ and mass parameter $\mu$ set to $\sqrt{0.1}$. Again we define $k=K/4\pi$. Note that the tiny deviations from the linear alignment of the chain-like $B=5,\,6^\star$ baby Skyrme configurations are purely numerical effects.}
\label{Fig_en_dens_V1_2}
\end{figure}

\begin{figure}[htb]
\centering
\begin{tabularx}{\textwidth}{XXX}
\epsfig{file=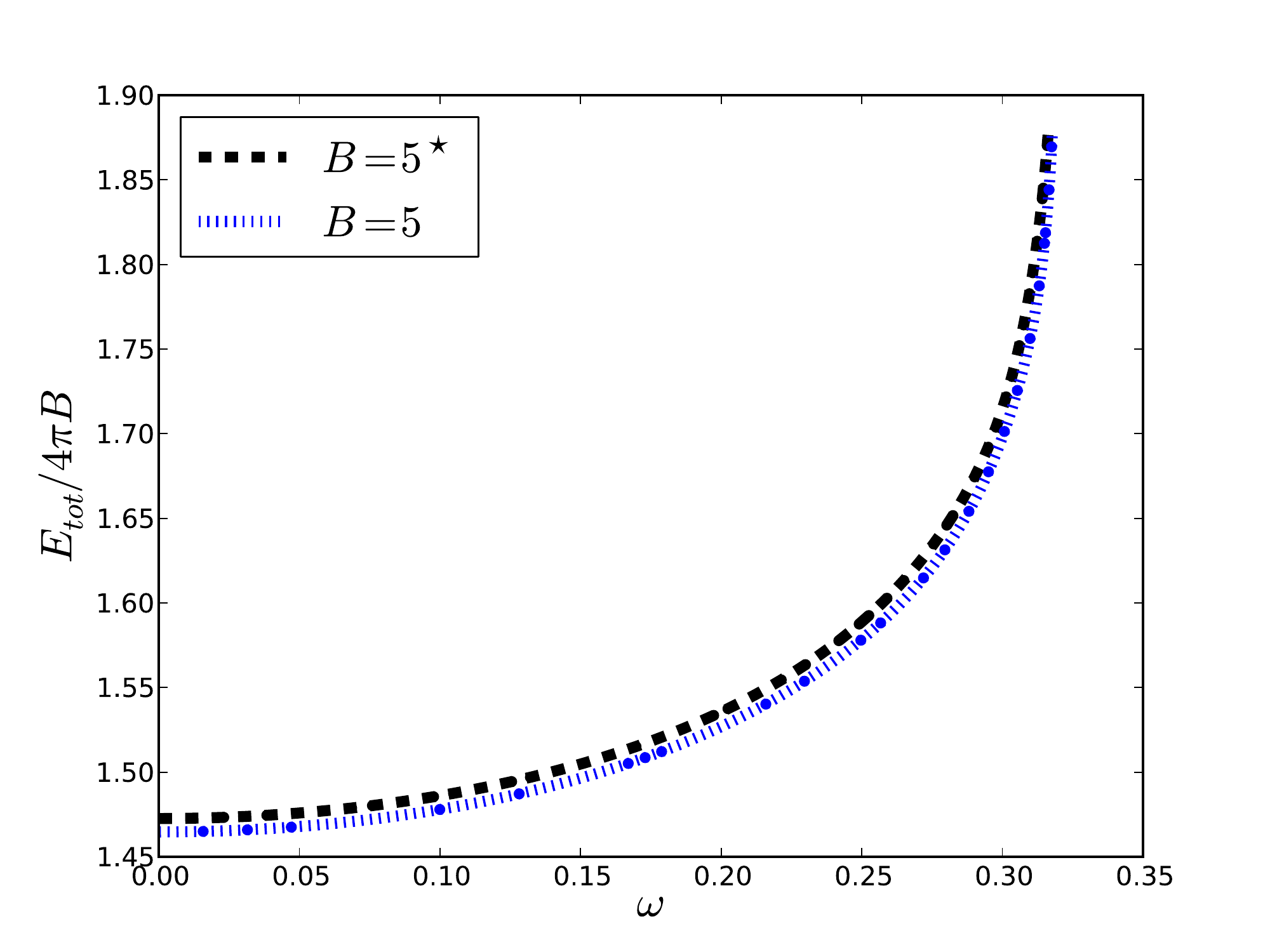,totalheight=4.0cm,clip=} &
\epsfig{file=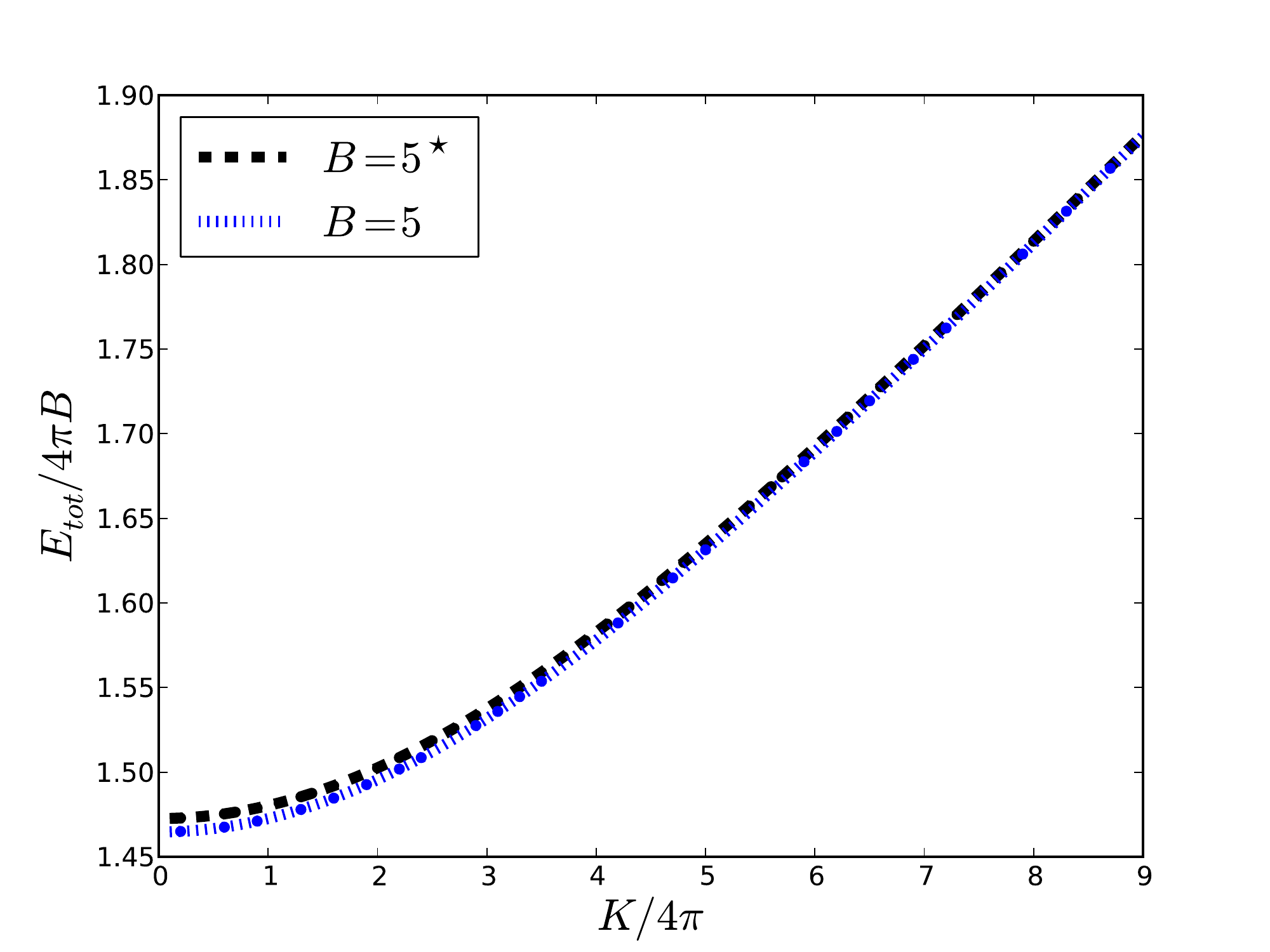,totalheight=4.0cm,clip=} &
\epsfig{file=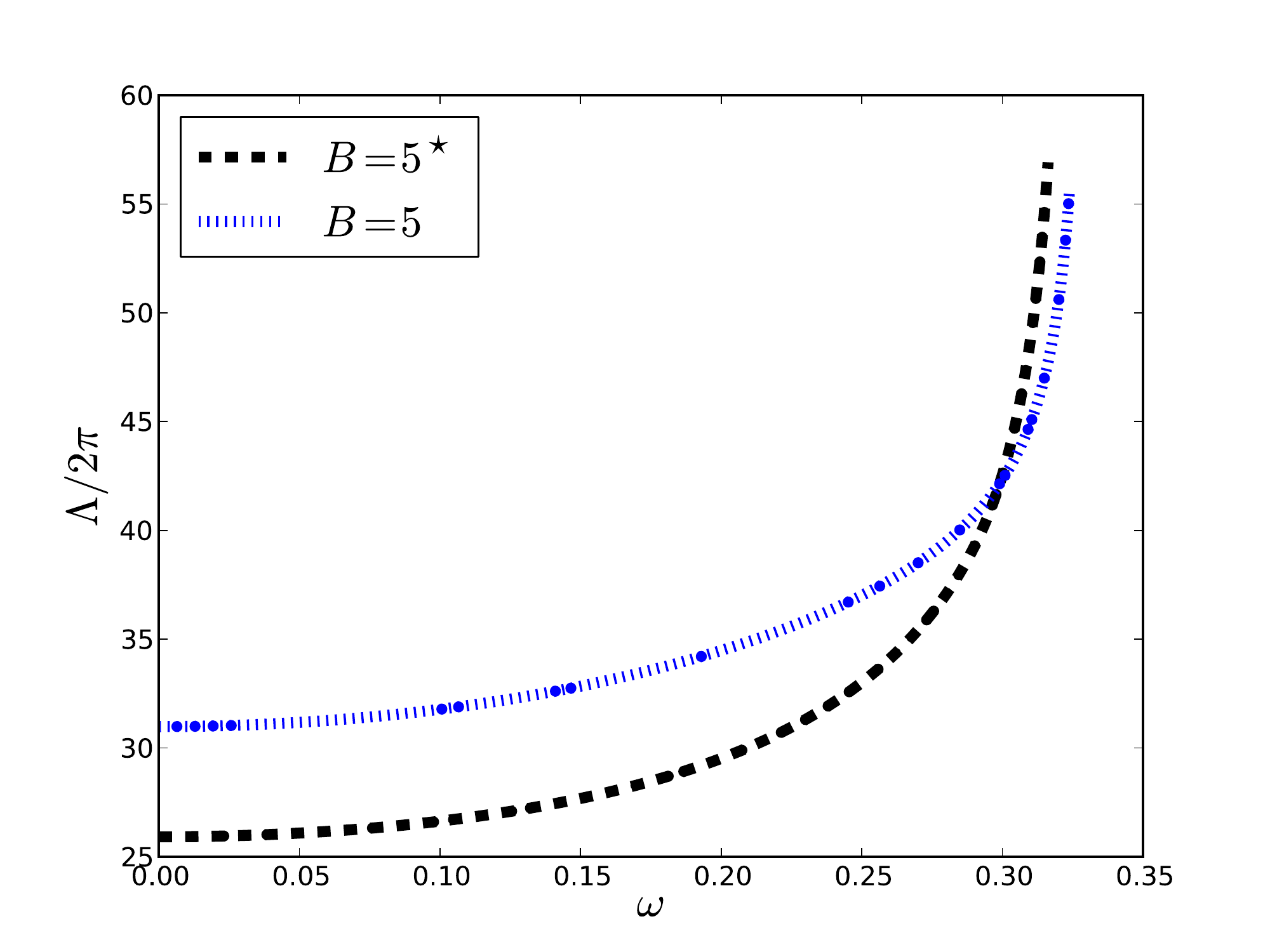,totalheight=4.0cm,clip=}
\end{tabularx}
\caption{Total energy $E_\text{tot}$ as a function of angular frequency $\omega$ and as a function of isospin $K$ for solitons in the standard baby Skyrme model with baryon number $B=5$ and mass value $\mu$ set to $\sqrt{0.1}$.}
\label{EwJ_B5_V1}
\end{figure}

\item $B=5$: With the mass parameter $\mu$ set to its standard value, the two different $B=5$ baby Skyrme configurations (5-chain solution and weakly bound $3+2$ solution) both split into 5 almost undistorted 1-solitons, see Fig.~\ref{Fig_en_dens_V1_2}. As already seen for the lower charge baby Skyrmion solutions, the deformations preserve the symmetries of the static, nonspinning Skyrmion solutions. Both Skyrme configurations are of very similiar energy and show, as a function of $\omega$ and $K$, almost identical energy curves; see Fig.~\ref{EwJ_B5_V1}.  

\begin{figure}[!htb]
\centering
\begin{tabularx}{\textwidth}{XXX}
\epsfig{file=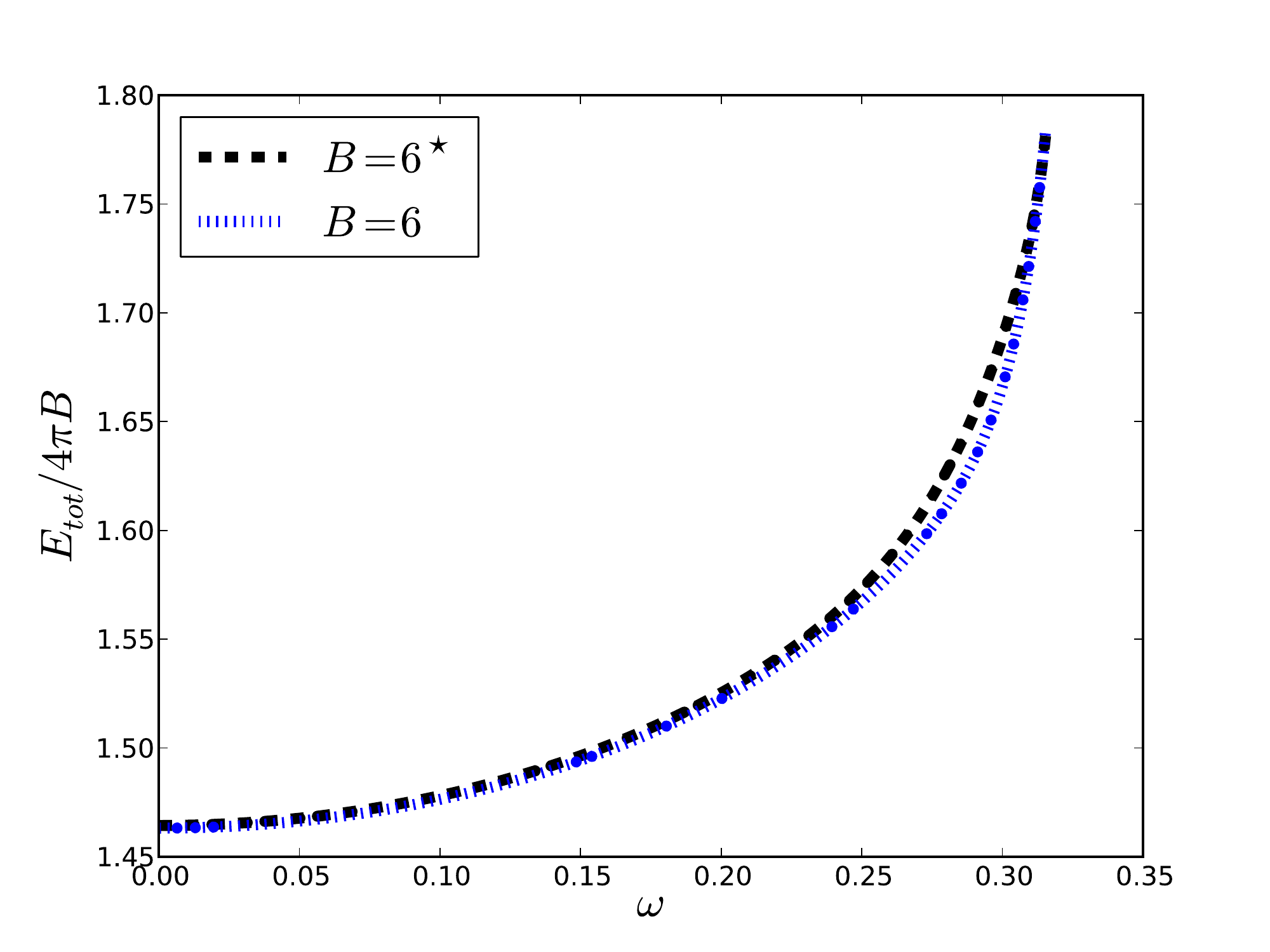,totalheight=4.0cm,clip=} &
\epsfig{file=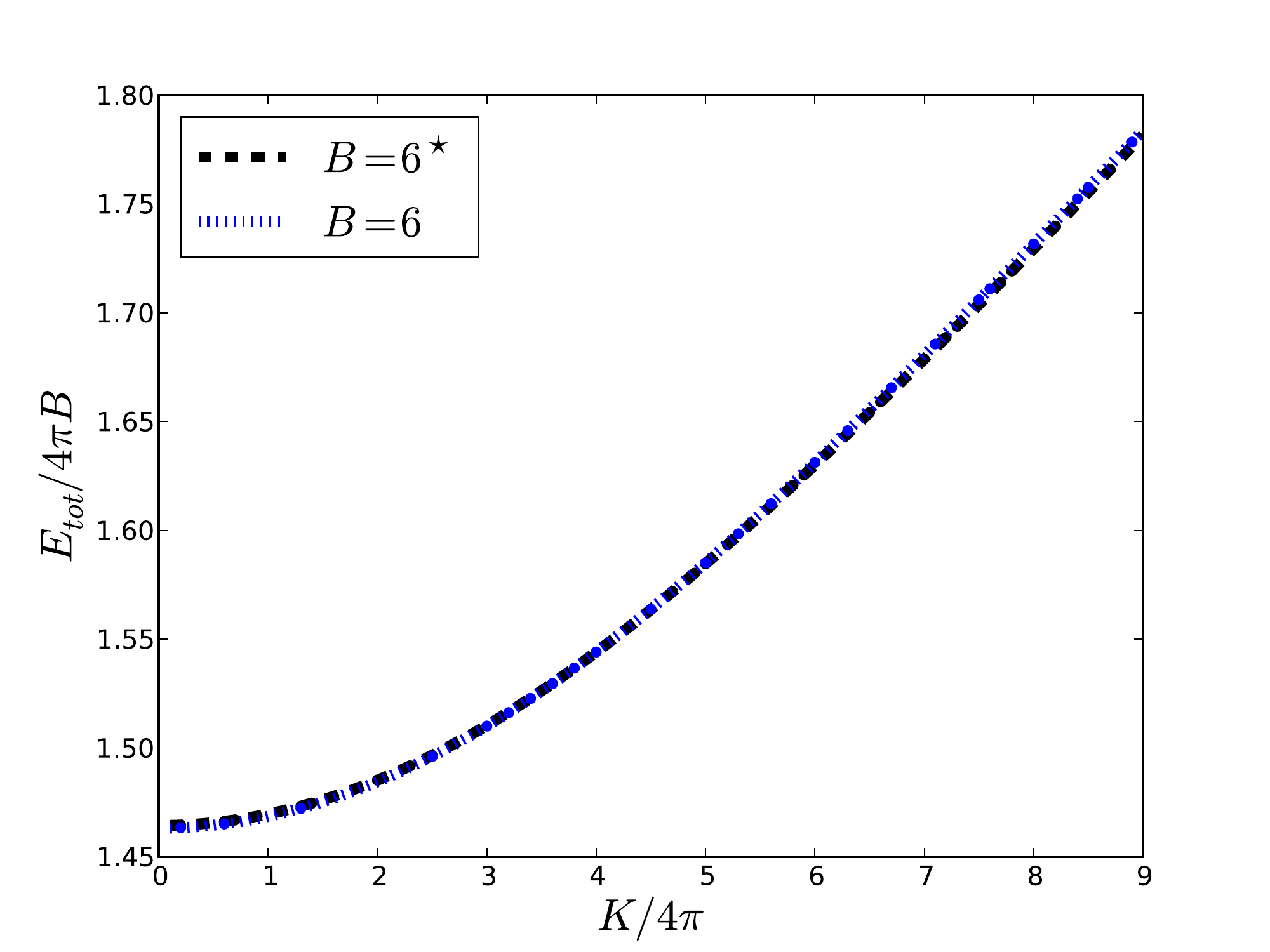,totalheight=4.0cm,clip=} &
\epsfig{file=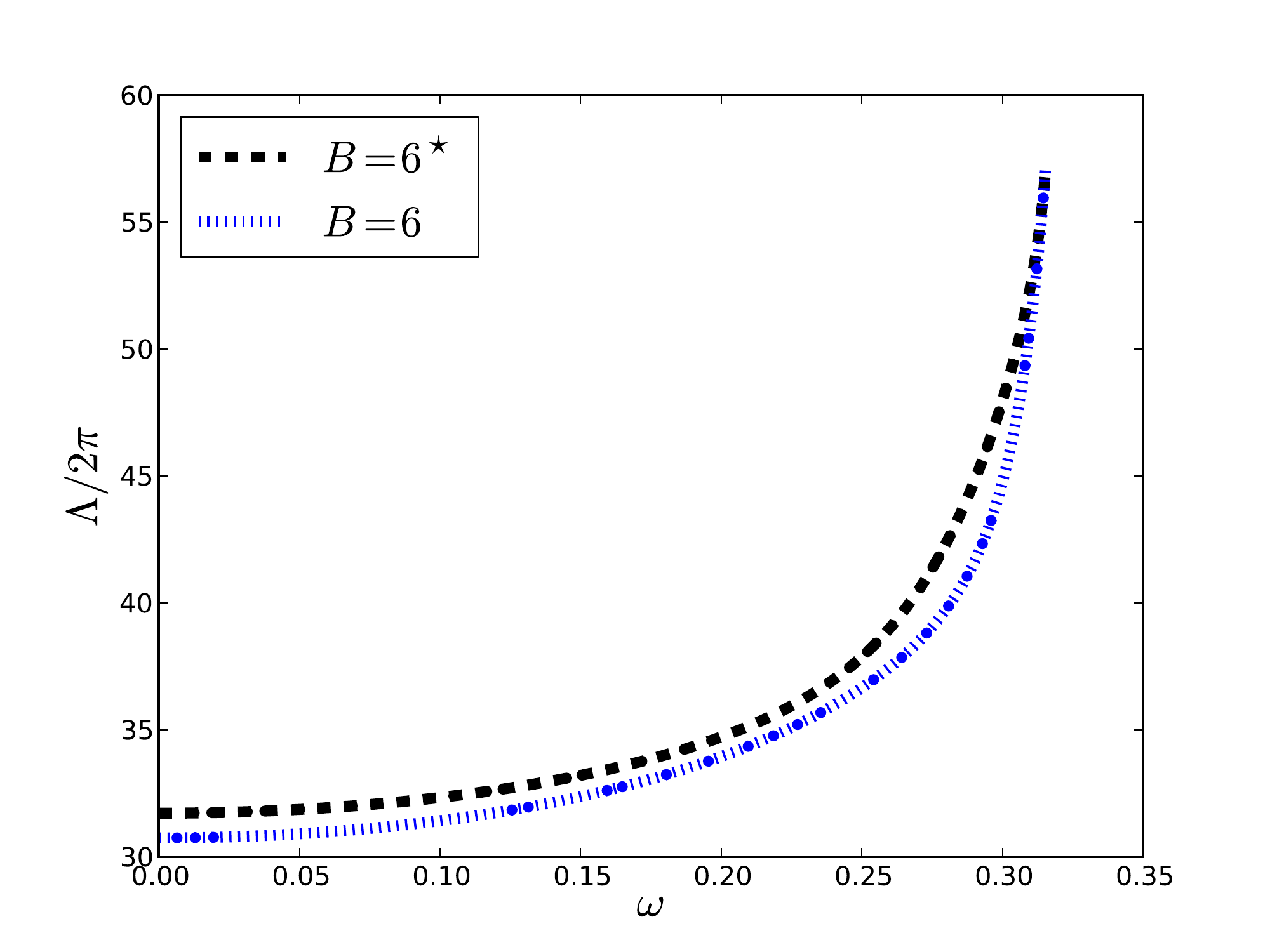,totalheight=4.0cm,clip=}
\end{tabularx}
\caption{Total energy $E_\text{tot}$ as a function of angular frequency $\omega$ and as a function of isospin $K$ for solitons in the standard baby Skyrme model with baryon number $B=6$ and mass value $\mu=\sqrt{0.1}$.}
\label{EwJ_B6_V1}
\end{figure}

\item $B=6$: The two (within the limits of our numerical accuracy) energy-degenerate 6-soliton configurations with $\mu=\sqrt{0.1}$ (6-chain solution and weakly bound $2+2+2$ solution) break into six single 1-baby Skyrmions. As above, the deformations do not break the symmetries of the nonspinning solutions; compare Fig.~\ref{Fig_en_dens_V1_2}. The energy degeneracy is not removed by isospinning the charge 6-solitons; see the energy curves given in Fig.~\ref{EwJ_B6_V1}. In particular, $E_{\text{tot}}(K)$ remains degenerate.   
\end{itemize}

\begin{figure}[!Ht]
\centering
\subfigure[\,$B=1,2$]{\includegraphics[totalheight=6.cm]{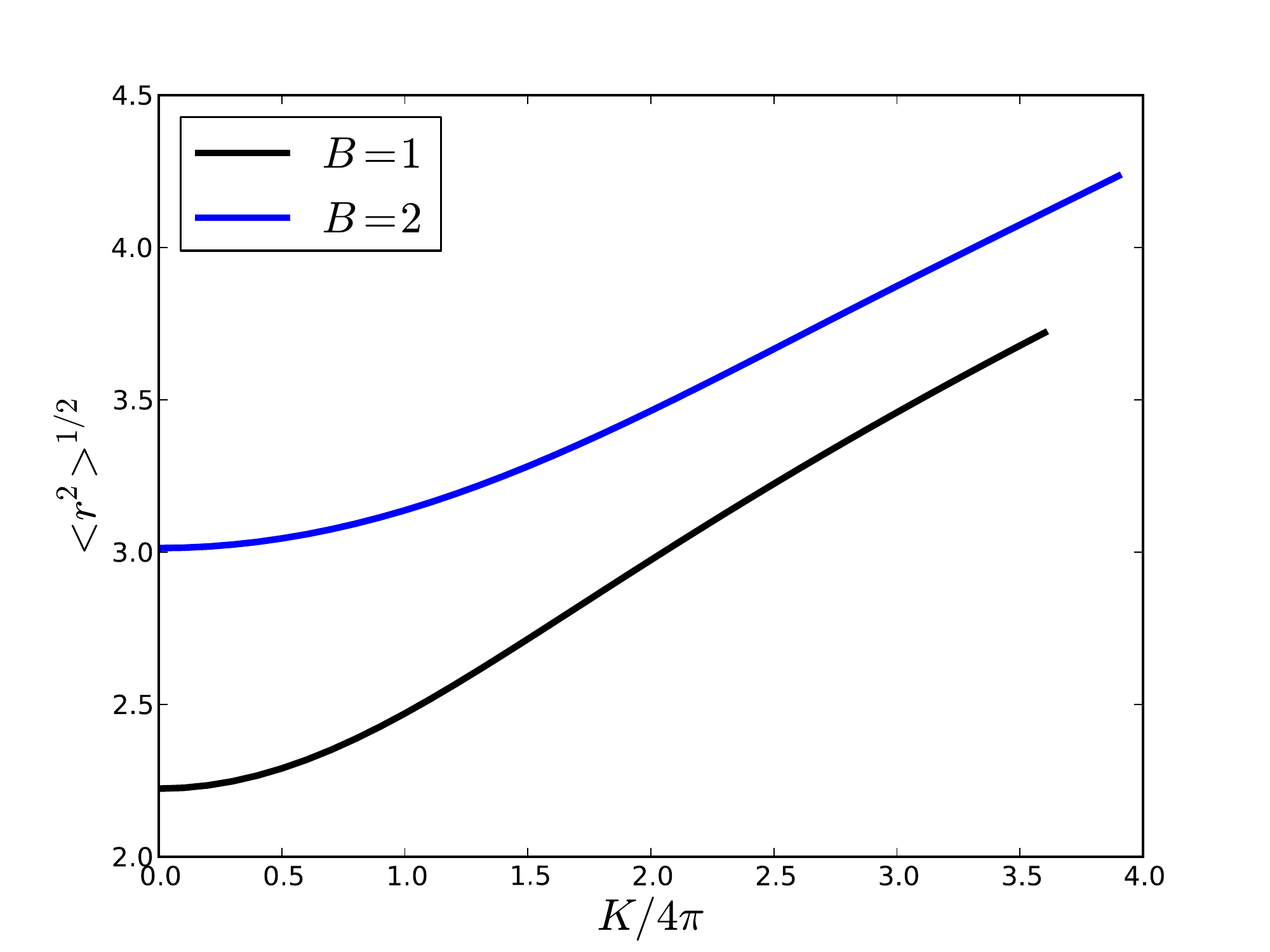}}
\subfigure[\,$B=1-6$]{\includegraphics[totalheight=6.cm]{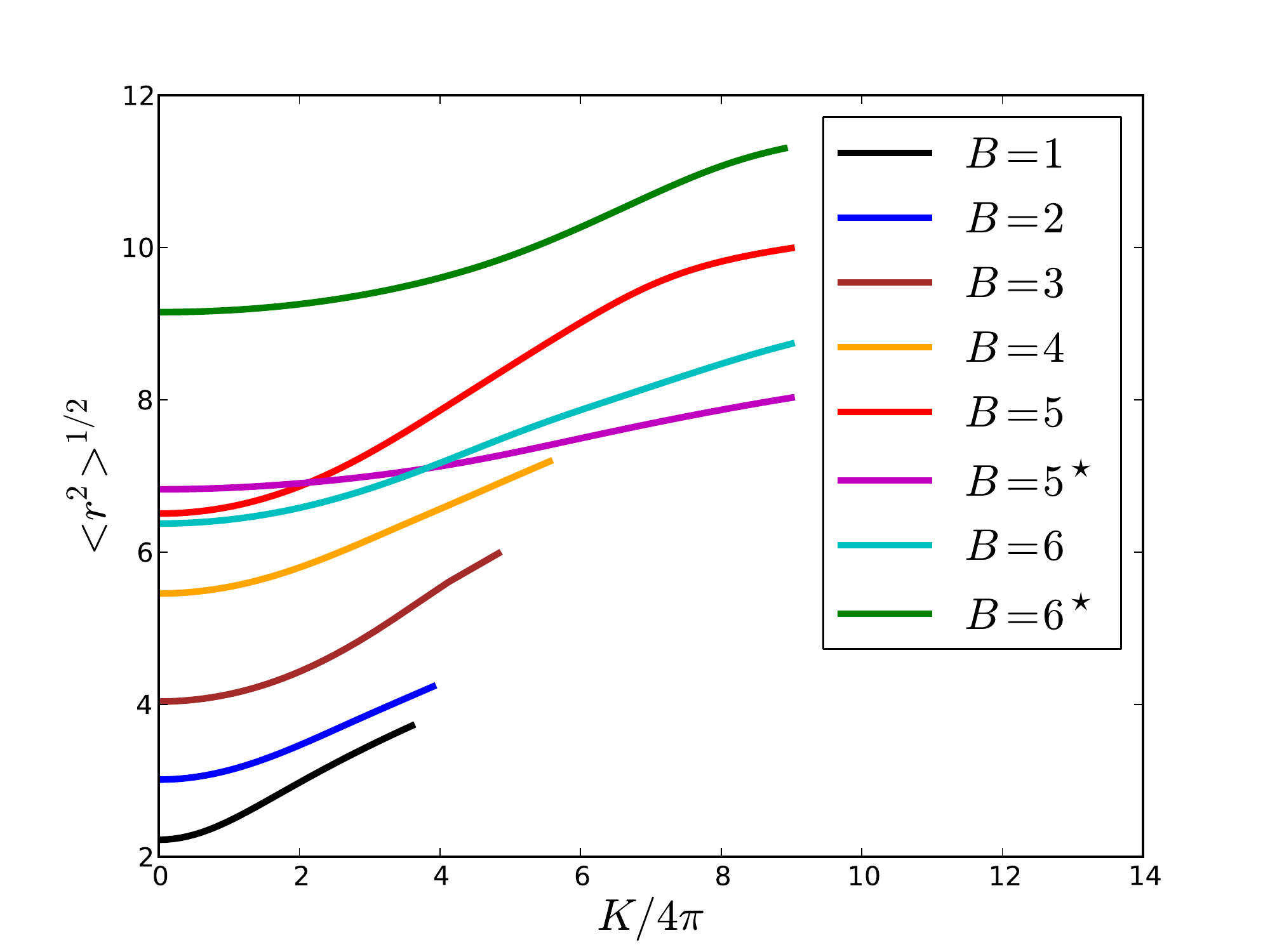}}
\caption{Mean Charge radii $<r^2\!>^{1/2}$ (\ref{Meancharge_radius}) for baby Skyrme solitons of topological charges $1\le B\le 6$ as a function of isospin $K$ and for mass value $\mu=\sqrt{0.1}$. }
\label{Fig_Meancharge}
\end{figure}

We display in Fig.~\ref{Fig_Meancharge} as a function of isospin $K$ the mean charge radii of $B=1-6$ baby Skyrmions (with $\mu=\sqrt{0.1}$) defined as the square root of the second moment of the topological charge density $B(x)$ (\ref{Baby_charge}),
\begin{equation}
<r^2>=\frac{\displaystyle\int\,r^2\,B(x)\,\text{d}^2x}{\displaystyle\int \,B(x)\,\text{d}^2x}\,.
\label{Meancharge_radius}
\end{equation}
The changes in the baby Skyrmions' shapes are reflected by the changes in slopes of the mean charge radius curves in Fig.~\ref{Fig_Meancharge}. We observe that for isospin values $K >1.04\times 4\pi$ the radius $<r^2\!>^{1/2}$ of the charge-1 solution grows approximately linear with $K$. For $B=2$  the  linear growth starts at higher angular momenta ($K\approx 1.51\times 4\pi$). These changes in slope are related to the rigid-body approximation only being a valid simplification for slowly isospinning Skyrme configurations, whereas for higher isospin values deformations due to centrifugal effects become increasingly important. Higher charge solutions can change their slopes several times. For example, the radius curve for the $5$-chain solution can be divided by its different slopes in three different regimes: For isospin values $K\le2.28\times 4\pi$ the charge-5 chain is made up of two $B=2$ tori weakly bound  together by a single $B=1$ baby Skyrmion. In the isospin range $2.28\times 4\pi<K\le 7\times 4\pi$ the chain is formed by the two tori moving further apart and the single $B=1$ constituent. Furthermore, an increase of $K$ results in five individual, linearly aligned  $B=1$ Skyrmions.  

\section{Conclusions}\label{Sec_Baby_Con}

We have performed full two-dimensional numerical relaxations of isospinning soliton solutions in the standard baby Skyrme model where the potential is given by the 2D analogue of the pion mass term of the full three-dimensional Skyrme model.

We find that completely analogous to the recent work on internally rotating soliton solutions \cite{Harland:2013uk,Battye:2013xf} in the 3D Skyrme-Faddeev model \cite{Faddeev:1975,Faddeev:1976pg}, there exist two types of critical frequencies: If the mass parameter $\mu$ is smaller than $1$, the isospinning configurations become unstable when the angular frequency $\omega$ approaches    $\mu$. If the mass parameter $\mu$ is taken to be larger than $1$, the energy of the isospinning solution becomes unbounded from below as $\omega$ increases above $1$. Hence, a stable isospinning soliton solution can only exist for frequencies  $\omega\le \text{min}(\mu,1)$. However, isospinning multi-Skyrmion solutions can break up into their constituent charge-1 Skyrmions before reaching this upper frequency limit. For $\mu$ sufficiently large there exists a third critical angular frequency value $\omega_3$ at which the total energy per unit charge is larger than the one of a single baby Skyrmion and the breakup into charge-1 baby Skyrmions is energetically favorable.

This pattern of critical behaviour has been unobserved in previous work \cite{Piette:1994mh} on isospinning charge-1 and -2 baby Skyrme solitons; mainly because the authors did not take into account deformations which break the rotational symmetry and only investigated relatively low mass values. Our numerical calculations clearly show that stable, rotationally symmetric Skyme solitons with mass parameter $\mu>1$ for angular frequencies $\omega>\omega_1=1$ are simply an artefact of the hedgehog approximation. Even for lower mass values ($\mu<1$) we find that the hedgehog ansatz can be a very poor approximation; the charge-2 baby Skyrmion solution can spontaneously break  its rotational symmetry when isospinning.

Further, we observe that for the conventional mass parameter choice ($\mu=\sqrt{0.1}$) the symmetries of the static, nonspinning soliton solutions are not significantly modified when isospin is added. This is in contrast to recent results on internally rotating soliton solutions \cite{Harland:2013uk,Battye:2013xf} in the 3D Skyrme-Faddeev model, where it was found that the model allows for transmutations, formation of new solution types and a rearrangement of the spectrum of minimal-energy solitons in a given topological sector when isospin is added. 

However, although  the soliton's geometrical shape is largely independent of the rotation frequency $\omega$, the soliton's size increases monotonically  with increasing $\omega$. In general, the rigid-body formula predicts for the solutions considered here total energies which for large angular momenta are roughly $1-10\%$ larger than those obtained for the deformed, isospinning solutions. Naturally, the accuracy of the rigid rotator approximation improves with increasing soliton mass and topological charge $B$.

\section*{Note added}
Note that similar results have been obtained independently by Alexey Halavanau and Yakov Shnir and have been reported in a very recent preprint \cite{Halavanau:2013vsa} which appeared shortly after ours. The authors use a rescaled version of the conventional baby Skyrme Lagrangian (\ref{Lag_Baby}) \cite{Piette:1994ug,Piette:1994mh}; the kinetic term differs from our notation by a factor of $2$. In particular, the mass parameter $\mu$ used in our article is related to the one ($\mu_{\text{HS}}$) used in \cite{Halavanau:2013vsa} by $\mu^2=\mu_{\text{HS}}^2/4$. We observe, for isospinning soliton solutions in the conventional baby Skyrme model, the same pattern of critical behaviour (see Figs.~1 and 2 in \cite{Halavanau:2013vsa}) and our results can be seen as complementary. Differences are the investigated mass range $\mu$ and the choice of initial conditions: Whereas we relax the absolute minima (especially nonrotationally symmetric configurations for $B>2$) at $\omega=0$ to find solutions for nonzero angular frequencies $\omega$, the authors in \cite{Halavanau:2013vsa} choose rotationally invariant ans\"atze as their starting configurations.

\section*{Acknowledgements}
We would like to acknowledge the use of the COSMOS Supercomputer in Cambridge. We thank David Foster, Juha J\"aykk\"a, Steffen Krusch and Paul Sutcliffe for useful discussions, and in particular, we are greatly indebted to Yakov Shnir for invaluable comments concerning the critical behavior of isospinning soliton solutions in baby Skyrme and Skyrme-Faddeev models. We thank Andrzej Wereszczy\'nski, Christoph Adam and Joaquin Sanchez-Guillen for pointing out references \cite{Gisiger:1996vb,Speight:2010sy,deInnocentis:2001ur,Adam:2010jr} and for valuable comments about the BPS limit of the standard Baby Skyrme model after a first version of our article appeared on the ArXiv. Some of the work of MH was undertaken at the SMSAS, University of Kent, financially supported by the UK Engineering and Physical Science Research Council (grant number EP/I034491/1).

\appendix*

\section{Isospinning charge-1 and -2 baby hedgehog solitons}\label{Appendix_1}

Previous numerical and analytical results \cite{Piette:1994mh,Gisiger:1996vb} on isospinning charge-1 and -2 baby Skyrmion solutions are largely based on the assumption that deformations are only happening within a rotationally symmetric hedgehog ansatz (\ref{Baby_hedge}). Consequently, previous work has been mainly concerned with the solution of Eq.~(\ref{ODE_hedgehog}). In this appendix, we briefly demonstrate that the pattern of critical behaviour observed for rotationally symmetric deforming Skyrmion solutions differs significantly from the one we found when allowing for arbitrary deformations.

\begin{figure}[!Ht]
\centering
\includegraphics[totalheight=7.cm]{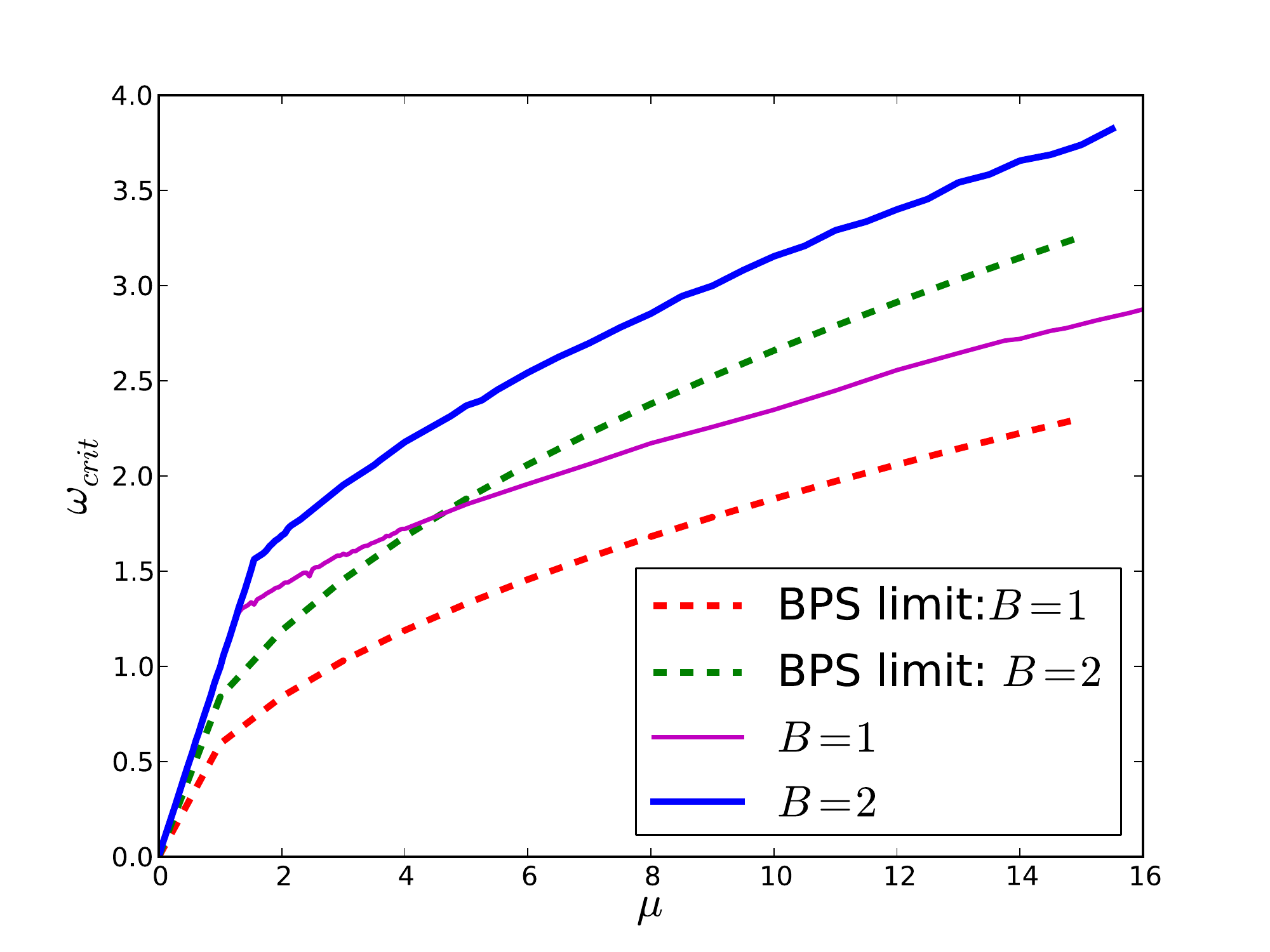}
\caption{Critical frequency $\omega_{crit}$ as a function of the mass parameter $\mu$ for isospinning $B=1$ and $B=2$ hedgehog soliton solutions in the full baby Skyrme model compared with the ones  in the BPS baby Skyrme model \cite{Gisiger:1996vb}. Note that solid lines represent the critical frequencies obtained by solving (\ref{ODE_hedgehog}) numerically, whereas dashed lines show the frequencies calculated analytically \cite{Gisiger:1996vb} in the infinite mass limit of the standard baby Skyrme model.}
\label{Mw_Crit_cut}
\end{figure}

For mass values $\mu\leq1$ the asymptotic behavior is governed by the $O(3)$ sigma model term and the potential term in the Skyrme Lagrangian (\ref{Lag_Baby}), whereas the Skyrme term is effectively removed. The linearized field equations give a critical angular frequency $\omega_{\text{crit}}=\omega_2=\mu$ and the spinning solitons are exponentially localized for $\omega<\omega_\text{crit}$ \cite{Piette:1994mh}. However, for larger $\mu$ ($\mu>1$) the Skyrme and potential term become increasingly dominant. For $\mu\rightarrow\infty$ the model is effectively described by the quartic (Skyrme) term and the potential. In this limit the  model is often referred to as  Bogomolny-Prasad-Sommerfield (BPS) baby Skyrme model \cite{Gisiger:1996vb,Adam:2010jr,Speight:2010sy} because its infinitely many exact static (multi)soliton solutions saturate the corresponding Bogomolny lower energy bound \cite{deInnocentis:2001ur,Adam:2010jr,Speight:2010sy}. Using various numerical methods (collocation \cite{Ascher:1979}, simple gradient flow evolution and Newton iteration \cite{petsc-web-page}) we solve (\ref{ODE_hedgehog}) for  isospinning charge-1 and -2 baby hedgehog solitons within the mass range $0<\mu\leq16$. The obtained critical angular frequencies $\omega_{\text{crit}}$ are shown in  Fig.~\ref{Mw_Crit_cut} as a function of the mass parameter $\mu$. For comparison, we also display in Fig.~\ref{Mw_Crit_cut} the analytically calculated \cite{Gisiger:1996vb} critical frequencies for charge-1 and -2 rotationally invariant Skyrme solitons in the $\mu\rightarrow\infty$ limit of the conventional baby Skyrme model. In this BPS limit the maximal rotation frequency has been calculated analytically  \cite{Gisiger:1996vb} to be given by  $\omega_\text{crit}=\sqrt{B\mu/2\sqrt{2}}$. 

We observe that the $\omega_{\text{crit}}(\mu)$ curves for isospinning hedgehog solutions in the full baby Skyrme model and in the BPS model are in qualitative agreement: In particular, the graphs show approximately the same asymptotic behavior and a crucially different behavior for low and higher mass values.

However, our full two-dimensional relaxation calculations in the standard baby Skyrme model reveal that isospinning soliton solutions are only stable up to angular frequencies $\omega\le \text{min}(\mu,1)$ and that the higher frequency value shown for $\mu>1$ in Fig.~\ref{Mw_Crit_cut} are purely an artefact of the hedgehog approximation.

\end{document}